\pdfoutput=1

\PassOptionsToPackage{hyperfootnotes=false,breaklinks=true}{hyperref}
\documentclass[usenatbib]{mnras}

\newcommand{\noopsort}[1]{}

\usepackage{newtxtext,newtxmath}
\usepackage[T1]{fontenc}
\usepackage{stfloats}

\makeatletter
\@namedef{figure*}{\@ifnextchar[{\mn@figstar}{\mn@figstar[tbp]}}
\def\mn@figstar[#1]{\def\baselinestretch{1}\st@rredfloattrue\let\@makecaption=\SFB@makefigurecaption\@dblfloat{figure}[#1]\centering\fstyle@figure}
\makeatother
\usepackage{graphicx}
\usepackage{amsmath}

\usepackage{amssymb}
\usepackage{tabularx}
\usepackage{booktabs}
\usepackage{placeins}
\usepackage[nolist,nohyperlinks]{acronym}

\newacro{HOD}{halo occupation distribution}
\newacro{SaUCE}{Sparse and Unbiased Clustering Estimator}

\title[Unbiased Galaxy Clustering from Fewer Merger Trees]{Less is More: Unbiased Galaxy Clustering from Fewer Merger Trees}
\author[Oscar J. Hickman et al.]{%
    Oscar J. Hickman,\thanks{E-mail: oscar.hickman@durham.ac.uk}
    Sownak Bose,
    Kai Wang,
    Carlton M. Baugh,
    Baojiu Li \\
    Institute for Computational Cosmology, Department of Physics, Science Laboratories, South Road, Durham, DH1 3LE, UK
}

\date{Accepted XXX. Received YYY; in original form ZZZ}
\pubyear{2026}

\begin{document}
\raggedbottom
\label{firstpage}
\pagerange{\pageref{firstpage}--\pageref{lastpage}}
\maketitle

\begin{abstract}
Galaxy clustering could strongly constrain galaxy formation models but
is rarely used, because the two-point correlation function (2PCF)
requires a complete galaxy catalogue at every point in parameter space.
Semi-analytic models (SAMs) process each halo merger tree independently,
so running a SAM on $m$ of $k$ disjoint partitions of the trees, each
spanning the full box, cuts the cost by $k/m$. Because a halo's
galaxies are kept or discarded together, one-halo pairs are diluted
less than two-halo pairs, and the naive estimate overstates small-scale
clustering by a factor $k/m$. We introduce the Sparse and Unbiased
Clustering Estimator (SaUCE), which removes this bias with weights set
by $m$ and $k$ alone. It has no free parameters, gives exactly unbiased
pair counts, and applies to any catalogue built from a random subset of
whole haloes and to any pair-count statistic. Running a SAM on 7 per
cent of the P-Millennium merger trees, 68 per cent of random selections
recover the full-box $\xi(r)$ at $z=0$ to within 0.05\,dex in each bin
from 0.03 to $30\,h^{-1}\,\mathrm{Mpc}$, at a density of
$10^{-2}\,h^{3}\,\mathrm{Mpc}^{-3}$: a fourteen-fold computational saving. The
required fraction rises with redshift and as the number density falls,
is the same for two model variants, and transfers between boxes through
the galaxy and pair counts. The scatter, derived in closed form, is
strongly correlated between bins, so a likelihood must use the full
covariance. A recipe for choosing the fraction makes the 2PCF
affordable for model calibration and simulation-based inference.
\end{abstract}

\begin{keywords}
methods: numerical -- methods: statistical -- galaxies: formation -- galaxies: haloes -- galaxies: statistics -- large-scale structure of Universe
\end{keywords}

\section{Introduction}

Galaxy clustering depends on how galaxies populate dark matter haloes. The physical processes of galaxy formation -- supernova feedback, feedback from active galactic nuclei (AGN), gas cooling, star formation and galaxy mergers -- shape this occupation, each changing the number and distribution of galaxies across haloes of different mass \citep{Baugh2006, Benson2010, Wechsler2018}. Clustering can constrain galaxy formation physics, and complements the one-point statistics, such as the luminosity function, on which model calibration has traditionally relied.

The advent of precision clustering measurements from surveys such as the Dark Energy Spectroscopic Instrument \citep{DESI2016} makes it even more timely to incorporate the 2PCF into model calibration.

Despite this, galaxy clustering is seldom used to calibrate physically motivated galaxy formation models. The principal obstacle is not evaluating the two-point correlation function (2PCF; \citealt{Peebles1980})
but the cost of producing the galaxy catalogue from which the clustering is measured.

Hydrodynamical simulations of cosmological volumes \citep[e.g.][]{Schaye:2023} are too expensive to run at every step of a parameter search, and are instead calibrated on smaller boxes \citep{Kugel:2023}, which can only probe the clustering over a narrow range of scales.

Semi-analytic models (SAMs) are cheaper by orders of magnitude, but still costly: a full galaxy catalogue for the P-Millennium simulation \citep{Baugh2019} requires processing the merger trees of more than 77 million haloes with masses above $2.12  \times 10^{9} h^{-1} M_{\rm{\odot}}$ at $z = 0$, and using the 2PCF in the calibration requires one such catalogue at every point in parameter space -- thousands of them for a Markov chain Monte Carlo (MCMC) search, and comparable numbers for simulation-based inference \citep{Cranmer2020}.

\citet{Henriques2013} made this affordable for one-point statistics by running the model on a small, representative sample of haloes, and \citet{vanDaalen2016} extended the approach to clustering. Calibrating the model of \citet{Guo2013}, they sampled haloes at a mass-dependent rate, using fewer than 0.04 per cent of all haloes in the Millennium I simulation, measured the halo occupation and satellite profiles of the sampled galaxies, and used these in a halo model \citep{Cooray2002} to predict the projected correlation function. Adding clustering to the MCMC narrowed degeneracies left open by the stellar mass function, tightening the supernova feedback parameters and constraining the timescale on which orphan satellites merge. The halo model, however, is itself an approximation: even applied to the full galaxy catalogue, their prediction departed from the directly measured clustering by up to $\sim$15 per cent outside their most massive bin, and the model's two-halo term took its halo clustering from linear theory and a fitted bias rather than from the sampled haloes.

\cite{Kaiser1986} showed that a sparsely sampled catalogue gives an unbiased estimate of the 2PCF, because random subsampling raises the shot noise without biasing the clustering signal. We apply this idea to merger trees. Because most SAMs process each tree independently, we can divide the trees of a cosmological volume into partitions, run the model on only some of them, and correct the correlation function for the fraction run, the partition fraction. Unlike Kaiser (1986), we sample haloes rather than galaxies: a halo's galaxies are kept or discarded together, so one-halo pairs, which survive whenever their halo is kept, are diluted less than two-halo pairs, which survive only if both haloes are kept. Without correction the small-scale clustering is overestimated, increasingly so as the partition fraction falls. We remove this bias exactly, with weights set only by the numbers of partitions run and available, so that, unlike the approach of \cite{vanDaalen2016}, nothing is tuned to a particular galaxy catalogue.

We introduce the \ac{SaUCE} and test it with the \texttt{GALFORM} SAM \citep{Cole2000,Bower2006,Lacey2016}. Section~\ref{sec:simulated-catalogues} describes the simulations, models and galaxy samples. Section~\ref{sec:clustering-formalism} derives the estimator, a correction with no free parameters, exact for the pair counts, that holds for any catalogue built from a random subset of whole haloes, and Section~\ref{sec:metrics} defines how we validate it. Section~\ref{sec:results} measures the partition fraction needed as a function of scale, redshift and number density; shows that it is the same for two \texttt{GALFORM} variants with different feedback but, at fixed number density, similar halo occupation distributions (HODs); shows how it transfers between simulations of different size; characterises the scatter and the bin-to-bin covariance a likelihood needs; and tests the correction in redshift space. Section~\ref{sec:conclusions} collects these into a practical recipe.

\section{Simulated galaxy catalogues}
\label{sec:simulated-catalogues}

We test \ac{SaUCE} with two variants of \texttt{GALFORM}  (Appendix~\ref{app:galaxy-formation-models}) run on three $N$-body simulations. \texttt{GALFORM} follows the baryonic components of galaxies within dark matter halo merger trees, each of which records the mergers and accretion by which a halo grew; this history sets the galaxy content of the halo and hence its contribution to the clustering. The two variants populate the same trees differently, which tests whether the scatter of SaUCE depends on galaxy formation physics.

The trees are extracted from the $N$-body simulations. SUBFIND subhaloes \citep{Springel2001} are grouped into DHalo haloes (\citealt{Jiang2014}; see also \citealt{Merson2013}) whose histories are strictly hierarchical, so no halo is shared between trees, a property \ac{SaUCE} requires (Section 3.5). Tree-building choices leave artefacts  that change the galaxy output \citep{Gomez:2022,ChandroGomez2025}, but the reference and sparse estimates use the same trees, so the comparison tests \ac{SaUCE} on whatever population those trees produce.

\subsection{$N$-body simulations}
\label{sec:N-body-simulations}

Table 1 lists the properties of the three simulations and the number of partitions, k, into which their merger trees are divided (Section~\ref{sec:merger-tree-partitioning}). The P-Millennium has by far the largest number of partitions, so its partition fraction can be set most finely, in steps of 1/1024.

Millennium~I and Millennium~II share $k = 64$ but differ in volume by a factor of $125$, so comparing them varies the box size at a fixed partition count (Section~\ref{sec:box-size-dependence}).

\begin{table}
\centering
\caption{Summary of $N$-body simulations used. $L$ is the box size, $k$ the number of merger-tree partitions, $N_{\rm p}$ the particle number and $M_{\rm p}$ the particle mass; the cosmology column lists the underlying parameters: Planck 2013 \citep{Planck2013} for the P-Millennium and approximately first-year WMAP \citep{WMAP1} for the Millennium simulations.
References: P-Millennium, \citet{Baugh2019}; Millennium~I, \citet{Springel2005}; Millennium~II, \citet{BoylanKolchin2009}.}
\label{tab:Simulations}
{\small\setlength{\tabcolsep}{3.5pt}
\begin{tabular}{lccccc}
\hline
Simulation & $L$ & $k$ & $N_{\rm p}$ & $M_{\rm p}$ & Cosmology \\
           & $(h^{-1}\,\mathrm{Mpc})$ & & & $(h^{-1}\,\mathrm{M}_\odot)$ & \\
\hline
P-Millennium & 542.2 & 1024 & $5040^3$ & $1.06\times10^{8}$ & Planck 2013 \\
Millennium I & 500 & 64 & $2160^3$ & $8.61\times10^{8}$ & WMAP1 \\
Millennium II & 100 & 64 & $2160^3$ & $6.88\times10^{6}$ & WMAP1 \\
\hline
\end{tabular}}

\end{table}

We run two variants of \texttt{GALFORM}, chosen because they populate the same merger trees differently: \texttt{L16} \citep{Lacey2016} and \texttt{GP14} \citep{GonzalezPerez2014}, which were calibrated for different purposes and differ in the parameters listed in Table~\ref{tab:galform}. Both models were originally calibrated on an $N$-body simulation of resolution similar to Millennium~I but with a different cosmology, and were recalibrated on the P-Millennium by \cite{Baugh2019}.
We run them on the Millennium I and Millennium II trees without further recalibration, so on those boxes they do not reproduce the calibration data in detail. This does not affect the tests here: SaUCE requires a galaxy population, not a realistic one, and the reference and sparse estimates are always drawn from the same population.

\subsection{Merger tree partitioning}
\label{sec:merger-tree-partitioning}
The merger trees of each simulation are divided between $k$ files, the partitions. Each tree is rooted on a halo at $z=0$, contains that halo's full progenitor history and lies in exactly one partition, so the partitions hold disjoint sets of haloes. We run \texttt{GALFORM} on each partition separately.

\begin{figure}
\centering
\begin{minipage}[t]{0.485\columnwidth}
\centering
\textbf{(a) Spatial sub-volume division}\par\smallskip
\includegraphics[width=\linewidth]{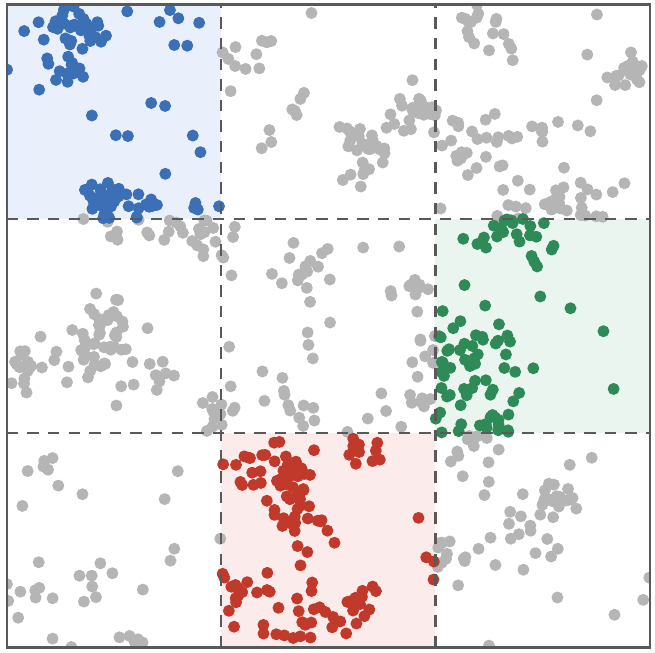}
\end{minipage}
\hfill
\begin{minipage}[t]{0.485\columnwidth}
\centering
\textbf{(b) Merger-tree sparse sampling}\par\smallskip
\includegraphics[width=\linewidth]{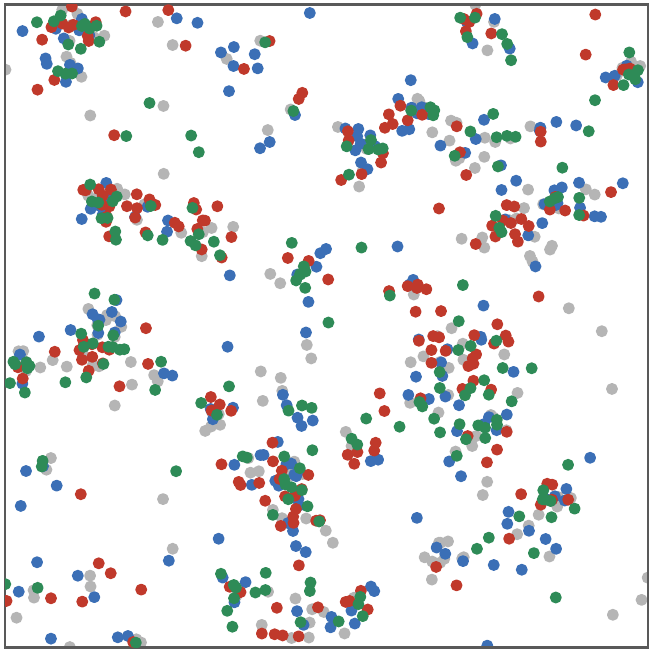}
\end{minipage}
\caption{Partition architecture. (a) Spatial sub-volume division, with haloes grouped by location; each coloured region is one contiguous sub-volume. (b) The merger-tree sparse-sampling or partitioning used here, with partitions (blue, red, green) assigned at random to $z=0$ merger trees rather than by spatial position; grey points are unselected haloes. Partition members are spread throughout the full simulation volume, preserving the spatial distribution.}
\label{fig:partitions}
\end{figure}

In cosmological simulations, how the haloes are distributed between the partitions is a free choice, illustrated in Fig.~\ref{fig:partitions}. The simulation volume is typically divided into spatial sub-volumes, so that the merger histories in one file all come from a single contiguous region many times smaller than the box.
For the simulations used here the tree-building code instead assigns each $z = 0$ halo, with its progenitors, to one of the $k$ files independently and with equal probability, so the haloes in a partition can come from anywhere in the volume and each partition spans the full box. Each partition therefore samples every scale the box resolves. This does not affect the bias of SaUCE (Section~3.5), but it keeps the scatter small, and the partition fractions of Section 5 depend on it. Because the assignment is random, the partitions are equal in size only on average; this spread adds to the variance of the estimate (Appendix~A1).

The random assignment applies to the $z = 0$ haloes. At $z > 0$ each halo is still selected with probability $m/k$, since it belongs to exactly one tree, but progenitors that share a $z = 0$ root are always selected together. This adds to the variance at $z > 0$ but not to the bias (Section 3.5).

\subsection{Galaxy sample definitions}
\label{sec:samples}

We define galaxy samples by number density rather than by stellar mass. A fixed stellar-mass threshold selects a different number density in every box, model and redshift, because each combination has its own stellar mass function; a number-density cut removes that confound, so that a comparison between boxes or between model variants isolates the quantity of interest. We rank galaxies by stellar mass and set the threshold that gives the target $\bar n$ in each full catalogue.

Three number-density samples underpin the convergence results. The galaxy counts below follow from $\bar n$ and the P-Millennium volume, so they are the same for either model and at every redshift; the stellar-mass thresholds are those of the \texttt{L16} model at $z = 0$, and vary with model and redshift:

{\raggedright
\begin{enumerate}
\item \textbf{$\bar n = 10^{-2}\,h^3\,\mathrm{Mpc}^{-3}$}:~$1.59\times10^{6}$ galaxies \mbox{($M_* > 10^{9.71}\,h^{-1}\,M_{\rm \odot}$)}.
\item \textbf{$\bar n = 10^{-3}\,h^3\,\mathrm{Mpc}^{-3}$}:~$1.59\times10^{5}$ galaxies \mbox{($M_* > 10^{10.64}\,h^{-1}\,M_{\rm \odot}$)}.
\item \textbf{$\bar n = 10^{-4}\,h^3\,\mathrm{Mpc}^{-3}$}:~$1.59\times10^{4}$ galaxies \mbox{($M_* > 10^{10.96}\,h^{-1}\,M_{\rm \odot}$)}. Shot-noise limited at small scales.
\end{enumerate}
\par}

Across all simulations, model variants and three redshifts, these densities correspond to stellar-mass thresholds spanning $M_* \approx 10^{9.4}$--$10^{11.5}\,h^{-1}\,M_\odot$. The comparisons between models and boxes (Sections~\ref{sec:model-robustness} and~\ref{sec:box-size-dependence}) use a fourth, denser sample, $\bar n = 0.0408\,h^3\,\mathrm{Mpc}^{-3}$, the density of $M_* > 10^{9}\,h^{-1}\,M_\odot$ in the P-Millennium \texttt{L16} catalogue at $z=0$ ($6.50\times10^{6}$ galaxies). We use a denser sample there because Millennium~II has $1/125$ of the volume of Millennium~I: at $\bar n = 10^{-2}\,h^3\,\mathrm{Mpc}^{-3}$ it would hold only $10^4$ galaxies.

We evaluate the convergence of the \ac{SaUCE} estimator at three redshifts, $z = 0$, $z \approx 0.5$ and $z \approx 1.5$, spanning the primary redshift coverage of surveys such as DESI \citep{DESI2016} and Euclid \citep{Laureijs2011}.

\subsection{Redshift-space catalogues}
\label{sec:rsd-catalogues}

We also consider the clustering in redshift space. \texttt{GALFORM} assigns each galaxy the peculiar velocity of its host halo or subhalo. We place a galaxy in redshift space by displacing it along one box axis, the line of sight, by $v_\parallel/(aH)$, where $v_\parallel$ is its line-of-sight peculiar velocity, $a$ the scale factor and $H$ the Hubble parameter at the snapshot redshift; displacements wrap at the periodic boundary.

\section{SaUCE estimator}
\label{sec:clustering-formalism}

We now derive the main result of the paper: an estimator that recovers
the full-box 2PCF from the galaxy catalogues of $m \le k$ partitions.

\subsection{Reference estimator}
\label{sec:reference-estimator}

We measure $\xi(r)$ from its definition as the excess probability, relative to a random distribution of points of the same mean density, of finding a galaxy pair at a given separation. For a separation bin centred on $r$,

\begin{equation}
\hat{\xi}(r) = \frac{DD(r)}{RR(r)} - 1, \qquad
RR(r) = \frac{1}{2}\,N_D (N_D - 1)\,\frac{\Delta V(r)}{V},
\label{eq:baseline-estimator}
\end{equation}

where $DD(r)$ is the number of galaxy pairs with separations in the bin, $N_D$ is the number of galaxies in the catalogue being counted, $V$ is the volume of the simulation box and $\Delta V(r)$ is the volume of the radial shell. $RR(r)$ is the number of pairs expected from $N_D$ points placed independently and uniformly in the box, and for a periodic box it is exact: with periodic boundaries every shell is complete for $r < L/2$, so there are no boundary or mask effects for a random catalogue to
measure, as there are in observational surveys \citep{LandySzalay1993}.

\subsection{Why a partition subset biases the estimate}
\label{sec:bias}

Random subsampling of galaxies scales $DD$ and $RR$ by the same factor,
and so adds shot noise but no bias \citep{Kaiser1986}.

Selecting $m$ of the $k$ partitions is different, because it samples haloes rather than galaxies: every galaxy in a halo is retained or discarded with its host. Whether a pair survives is then decided by how many partitions it occupies, so what we need are the probabilities that one, and that two, specified partitions are among those drawn. The $m$ are drawn uniformly from the $k$ without replacement, and we call one such draw a selection. Any one partition is drawn with probability $m/k$, and once it has been, a second specified partition must be one of the remaining $m-1$ taken from the remaining $k-1$:

\begin{equation}
p_1 = \frac{m}{k}, \qquad
p_2 = \frac{m(m-1)}{k(k-1)}.
\label{eq:survival}
\end{equation}

A pair confined to one partition survives with probability $p_1$, and a
pair spread across two survives only if both are drawn, with
probability $p_2$.

Galaxy pairs divide into one-halo pairs, whose two galaxies share a
halo, and two-halo pairs, whose galaxies occupy different haloes
\citep[e.g.][]{Cooray2002}. How a sample divides between the two
is set by its HOD \citep{Berlind2003, Zheng2005}, which differs
between galaxy formation models and between samples. No halo is split
across partitions (Section~\ref{sec:merger-tree-partitioning}), so both galaxies of a one-halo pair lie in one partition and the pair survives with probability $p_1$. Two-halo pairs generally have their haloes in different partitions, and survive with probability $p_2$: at $z = 0$ two distinct haloes share a partition with probability only $1/k$, although at $z > 0$ progenitors of a common root always do (Section~3.5).

The naive estimate normalises the combined catalogue by its own $RR$, which scales on average as the square of the galaxy count and hence as $(m/k)^2$. Two-halo pairs in different partitions are retained at $p_2 = (m/k)^2\,(1-1/m)/(1-1/k)$, so they are understated by a factor of about $(m-1)/m$; the fraction $1/k$ that share a partition are retained at $p_1$ and overstated by $k/m$, so at $z = 0$ two-halo pairs as a whole are retained, on average, at exactly $(m/k)^2$. One-halo pairs, all retained at $p_1 = m/k$, are overstated by a factor $k/m$. For $k = 1024$ and $m = 8$ this is a factor of 128.

\subsection{Derivation of the weights}
\label{sec:sauce-derivation}

What sets a pair's survival probability is not whether it is a one-halo
or a two-halo pair, but whether its galaxies lie in one partition or
two. We therefore weight pairs by partition, with the corrected
pair count:

\begin{equation}
DD(r) = w_{\rm auto}\, DD_{\rm auto}(r) + w_{\rm cross}\, DD_{\rm cross}(r),
\label{eq:corrected-count}
\end{equation}

where $DD_{\rm auto}$ is the number of pairs whose galaxies lie in the
same selected partition and $DD_{\rm cross}$ the number whose galaxies
lie in two different selected partitions. Because the classification is
by partition rather than by halo, it can be read directly off the
catalogue, and no approximation enters. This weighted count, not the
raw sum $DD_{\rm auto} + DD_{\rm cross}$ of the naive estimate, is the
$DD(r)$ that enters equation~\eqref{eq:baseline-estimator}.

Write $DD_{\rm auto}^{(k)}$ and $DD_{\rm cross}^{(k)}$ for the same two counts in the full $k$-partition catalogue; together they give the full-box pair count. Every pair in $DD_{\rm auto}^{(k)}$ survives with probability $p_1$ and every pair in $DD_{\rm cross}^{(k)}$ with probability $p_2$. Writing $\mathbb{E}$ for the average over the random choice of partitions, at fixed full-box catalogue, summing over pairs gives, exactly,

\begin{align}
\mathbb{E}\!\left[DD_{\rm auto}\right]  &= p_1\, DD_{\rm auto}^{(k)}, \label{eq:exact-auto}\\
\mathbb{E}\!\left[DD_{\rm cross}\right] &= p_2\, DD_{\rm cross}^{(k)}. \label{eq:exact-cross}
\end{align}

On average, the selected catalogue's $RR$ is smaller than the full-box
$RR$ by $(m/k)^2$, so the corrected count must on average be smaller
than the full-box count by the same factor:

\begin{equation}
\begin{split}
w_{\rm auto}\, p_1\, DD_{\rm auto}^{(k)} &+ w_{\rm cross}\, p_2\, DD_{\rm cross}^{(k)} \\
&= \frac{m^2}{k^2}\left(DD_{\rm auto}^{(k)} + DD_{\rm cross}^{(k)}\right).
\end{split}
\end{equation}

One pair of weights must serve every separation bin, and $DD_{\rm auto}^{(k)}$ and $DD_{\rm cross}^{(k)}$ stand in a different ratio in each bin, so the two classes must be matched separately. Each class is divided by its survival probability and multiplied by $(m/k)^2$:

\begin{equation}
w_{\rm auto} = \frac{(m/k)^2}{p_1} = \frac{m}{k}, \qquad
w_{\rm cross} = \frac{(m/k)^2}{p_2} = \frac{m(k-1)}{k(m-1)}.
\label{eq:weights}
\end{equation}

Up to the common factor $(m/k)^2$, which matches the normalisation of the selected catalogue, these are Horvitz--Thompson weights \citep{Horvitz1952}: each pair is weighted by the inverse of its probability of being selected, so the weighted count is an unbiased estimate of the full-box count. The partitions play the role of clusters in single-stage cluster sampling \citep{Cochran1977}, and a pair's selection probability is the probability that the one or two partitions it occupies are all drawn. The same idea underlies the pairwise inverse-probability weights that correct for galaxies left unobserved by fibre assignment in redshift surveys \citep{BianchiPercival2017}; here the probabilities are known exactly, rather than estimated from many realisations of the assignment.

Equations~\eqref{eq:exact-auto} and \eqref{eq:exact-cross} involve no approximation, so the corrected pair count is exactly unbiased (Appendix~\ref{app:closed-form-variance}). $\hat{\xi}$ itself is a ratio: $RR$ is built from the selected galaxy count, which varies from one selection to another, so $\hat{\xi}$ is unbiased only to leading order; Appendix~\ref{app:mean-vs-median} finds its mean offset from the reference consistent with zero.

Setting $m = k$ gives $w_{\rm auto} = w_{\rm cross} = 1$ and recovers the full catalogue. The smallest usable configuration is $m = 2$, the fewest partitions for which cross pairs exist.

Substituting equation~\eqref{eq:weights} into equations~\eqref{eq:corrected-count} and~\eqref{eq:baseline-estimator} gives the \ac{SaUCE} estimator,

\begin{equation}
\hat{\xi}(r) = \frac{1}{RR(r)}\left[\frac{m}{k}\, DD_{\rm auto} + \frac{m(k-1)}{k(m-1)}\, DD_{\rm cross}\right] - 1,
\end{equation}
with $RR(r)$ built from the galaxy count $N_D$ of the combined $m$-partition catalogue.

\subsection{Redshift space}
\label{sec:redshift-space}

Converting redshift to radial distance under the assumption of pure Hubble flow (Section~\ref{sec:rsd-catalogues}) displaces galaxies along the line of sight by their peculiar velocities, making the clustering anisotropic \citep{Kaiser1987}. We bin pairs in the redshift-space separation $s$, measured from the displaced positions, and in $\mu$, the cosine of the angle between the pair separation and the line of sight. The periodic-box estimator generalises directly,
\begin{equation}
\hat{\xi}(s,\mu) = \frac{DD(s,\mu)}{RR(s,\mu)} - 1,
\end{equation}
where $RR(s,\mu)$ is given by equation~\eqref{eq:baseline-estimator} with the volume of the $(s,\mu)$ bin in place of $\Delta V(r)$. The anisotropy is summarised by the multipole moments \citep[e.g.][]{Sanchez2016}

\begin{equation}
\hat{\xi}_\ell(s) = \frac{2\ell+1}{2} \int_{-1}^{1} \hat{\xi}(s,\mu)\, L_\ell(\mu)\, \mathrm{d}\mu,
\label{eq:multipoles}
\end{equation}

where $L_\ell$ is the Legendre polynomial of order $\ell$; the monopole ($\ell = 0$) and quadrupole ($\ell = 2$) carry most of the signal.

The correction is unchanged. The derivation of Section~\ref{sec:sauce-derivation} is made bin by bin and uses only each galaxy's partition label, which a line-of-sight displacement does not change, so the same weights apply to $DD(s,\mu)$, and indeed to any binning of the pairs, including the projected correlation function $w_{\rm p}(r_{\rm p})$. Each multipole is a linear functional of $\hat{\xi}(s,\mu)$, so the multipoles inherit the same leading-order unbiasedness.

\subsection{Assumptions and validity}
\label{sec:scope-validity}

The weights require only that whole partitions are selected, $m$ of the
$k$ drawn uniformly without replacement, and that whether a galaxy
enters the sample does not depend on which partitions are drawn.
In our tests the stellar-mass threshold is fixed from the full
catalogue (Section~4.1); in a calibration run it must instead be set
from the sparse catalogue itself, and then varies slightly between
selections. Galaxies near the threshold then enter some selections and
not others, which in our tests shifts the mean $\xi$ by at most 1 per cent and barely changes the scatter.

Two further conditions ensure that the sparse catalogue contains the
same galaxies as the full one. The galaxy formation model must process
each merger tree independently, so that running it on a subset of
trees reproduces those trees' galaxies; this holds for standard
semi-analytic models but fails for processes that couple trees, such as
a non-local treatment of reionisation. And no tree may be split between
partitions, since the model needs a halo's full history to populate it;
the weights themselves use only each galaxy's partition label.

The weights place no condition on how trees are assigned to
partitions, since equations~(4) and~(5) hold however the full-box pairs
divide between $DD^{(k)}_{\rm auto}$ and $DD^{(k)}_{\rm cross}$. Two
consequences follow. Archives that store trees in files by spatial
region can use those files directly as partitions: the estimate remains
unbiased, but its variance is larger, because compact partitions sample
the box unevenly (Fig.~1), so the partition fractions of Section~5
should not be assumed to carry over. And because our trees are
partitioned at $z = 0$, progenitors that share a root always lie in the
same partition, so at $z > 0$ partition membership is correlated with
pair separation. This changes how pairs divide between the two classes,
and hence the variance, but not the bias.

The correction acts only on $DD$. We normalise analytically because the
box is periodic, which leaves the choice of partitions as the only
source of scatter (Appendix~A), but the correction works equally with
$RR$ counted from a random catalogue, as in the estimator of
\citet{LandySzalay1993} for a survey geometry.

\section{Validation}
\label{sec:metrics}

Every result in Section~\ref{sec:results} compares a \ac{SaUCE} estimate built from $m$ partitions with a reference built from all $k$; we call $f = m/k$ the partition fraction. Each selection (Section~\ref{sec:bias}) corresponds to the first $m$ partitions of a random permutation of the $k$, so selections at fixed $m$ are independent, while a larger $f$ contains every partition used at a smaller one, and results at neighbouring $f$ are correlated. Unless stated otherwise, each configuration is measured over $105$ permutations; the convergence thresholds use $10{,}500$ (Section~\ref{sec:acceptance-criteria}).

\subsection{Reference}
\label{sec:ref-definitions}

The reference $\xi_{\rm ref}(r)$ is the correlation function measured from all $k$ partitions, from equation~\eqref{eq:baseline-estimator}. All pair counts, for the reference and for every selection, are tabulated with \texttt{Corrfunc} \citep{sinha2020corrfunc}. Each result uses a subset of the test cases of Sections~\ref{sec:N-body-simulations} and~\ref{sec:samples} and states which; within a case, the same stellar-mass threshold is applied to the reference and to every selection.

\ac{SaUCE} estimates the full-box correlation function of a given $N$-body realisation. The reference is the estimate made using all of the partitions, so cosmic variance of the realisation does not enter the comparison; an application must add it separately (Section~\ref{sec:covariance}).

\subsection{Separation bins}
\label{sec:bins}

We count pairs in logarithmic bins of width 0.1\,dex. The two ends of the range are not free, being set by the resolution of the simulation and by the size of the box; the width is a resolution choice.

The bin range is set by the simulation. The P-Millennium resolves haloes down to $M_{\rm h} = 2.12\times10^{9}\,h^{-1}\,\mathrm{M}_\odot$, the mass of $20$ particles \citep{Baugh2019}, whose comoving $r_{200}$ (at $200$ times the mean density) is $0.031\,h^{-1}\,\mathrm{Mpc}$ and is independent of redshift in comoving units. Pairs closer than this increasingly involve satellites whose subhaloes the simulation does not resolve, so the separation is set by \texttt{GALFORM}'s orbital assignment rather than by the $N$-body calculation. We therefore report $r \geq 0.03\,h^{-1}\,\mathrm{Mpc}$, the nearest bin centre on the grid to that scale. This floor comes from resolution: at $\bar n = 10^{-2}\,h^3\,\mathrm{Mpc}^{-3}$ every bin down to $r = 0.003\,h^{-1}\,\mathrm{Mpc}$ holds more than $2000$ reference pairs, ten times closer in than we report.

The outer edge is set by the box. Clustering predictions are conventionally restricted to a tenth of the box side
\citep[e.g.][]{Orsi:2008}, which is 54\,$h^{-1}$\,Mpc for the P-Millennium. We report $r \le 30\,h^{-1}$\,Mpc, comfortably inside this limit and the largest separation at which $\xi_{\rm ref}$ is positive in every sample and redshift tested.

The scatter fits of Section~\ref{sec:variance-recipe}, the covariance of Section~\ref{sec:covariance} and the tests of Appendix~\ref{app:variance-skew} use a coarser grid: 30 bins of width 0.148\,dex
with limits of 0.01 and 271\,$h^{-1}$\,Mpc; in Section~\ref{sec:variance-recipe}, Millennium~I and~II use 30 bins of 0.123\,dex, from 0.01 to 50 and from 0.002 to 10\,$h^{-1}$\,Mpc respectively.
The fits and Appendix~\ref{app:variance-skew} keep the 25 bins with centres below 50\,$h^{-1}$\,Mpc, and the covariance keeps the first 27, spanning 0.01--98\,$h^{-1}$\,Mpc. These ranges are wider than the one above because they characterise the scatter of the estimator, not its accuracy.

\subsection{Accuracy metrics}
\label{sec:accuracy-metrics}

We measure accuracy on the ratio of a selection's estimate to the reference,

\begin{equation}
R(r) = \frac{\hat\xi(r)}{\xi_{\rm ref}(r)},
\label{eq:rho}
\end{equation}
over the ensemble of independent selections at fixed $m$. We report two quantities:

\begin{enumerate}
\item the \textbf{bias}, $\langle R \rangle - 1$, the mean over selections. The mean is the statistic in which \ac{SaUCE} is unbiased; a median of $R$, or a mean of $\log_{10}R$, is offset from zero even for an exactly unbiased estimator (Appendix~\ref{app:variance-skew}); \item the \textbf{$1\sigma$ band}, the 16th--84th percentile range of $\log_{10}R$, in dex. This is the scatter a user running a single selection faces. Its half-extent is the larger of the distances of the 16th and 84th percentiles from zero, so an offset or asymmetric band is judged by its worse side. The distribution of $R$ is close to lognormal (Appendix~\ref{app:variance-skew}), so the width of the band follows from the variance, which Appendix~\ref{app:closed-form-variance} predicts in closed form and Section~\ref{sec:variance-recipe} fits.
\end{enumerate}
Results are sometimes broken down by clustering regime rather than bin by bin, as diagnostic columns in Table~\ref{tab:convergence}. Those breakdowns use three ranges: one-halo, $r \in [0.3, 1]$; transition, $r \in [1, 5]$; and two-halo, $r \in [5, 30]\,h^{-1}\,\mathrm{Mpc}$, holding six, seven and eight bins respectively on this grid.

\begin{figure*}
	\centering
	\includegraphics[width=0.33\textwidth]{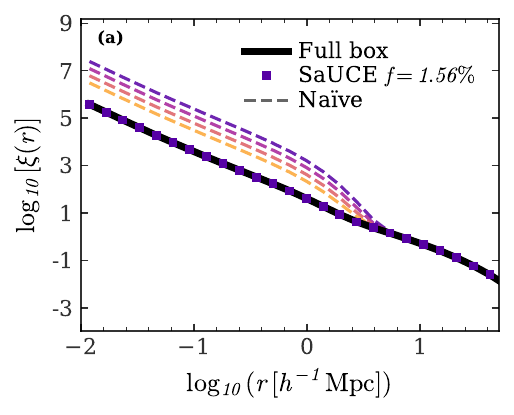}
	\includegraphics[width=0.33\textwidth]{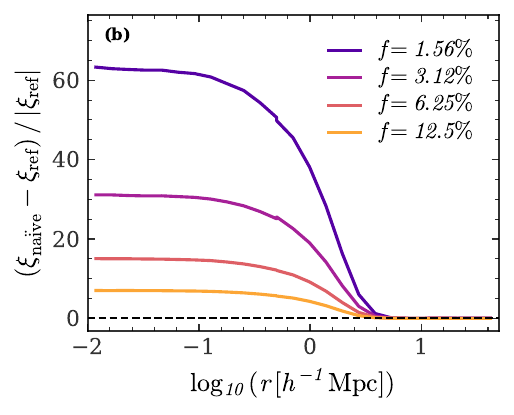}
	\includegraphics[width=0.33\textwidth]{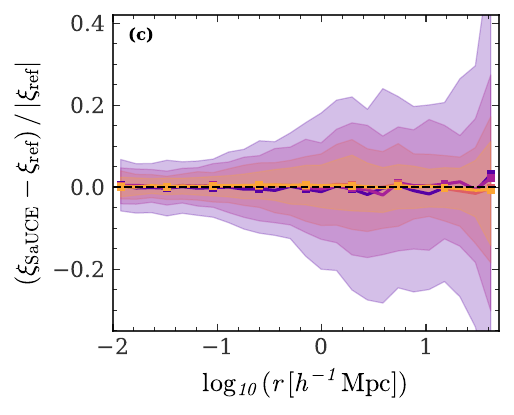}
	\caption{The naive and \ac{SaUCE} estimates of $\xi(r)$ against the reference, for $\bar{n} = 10^{-2}\,h^3\,\mathrm{Mpc}^{-3}$ in the \texttt{L16} model on the P-Millennium at $z = 0$. (a)~$\xi(r)$ from the naive and \ac{SaUCE} estimators at $f = 1.56$ per cent, against the reference (all $k = 1024$ partitions; solid black), from which the \ac{SaUCE} estimate is indistinguishable on this axis. (b)~Fractional residual of the naive estimator, $(\xi_{\rm naive} - \xi_{\rm ref})/|\xi_{\rm ref}|$. On small scales, where almost all pairs are one-halo pairs, it reaches $k/m - 1$ (Section~\ref{sec:bias});
    it falls to zero on large scales, where almost all pairs are two-halo pairs. (c)~Fractional residual of the \ac{SaUCE} estimator. Lines show the mean over the selections available at each $f$, between $192$ and $415$ of them; shaded bands span the 16th--84th percentiles, the scatter of a single selection. The panels extend below the reported range, $r \geq 0.03\,h^{-1}\,\mathrm{Mpc}$, to show the naive bias reaching its plateau. Note the different $y$-axis ranges of panels (b) and (c).}
	\label{fig:method-improvement}
\end{figure*}

\subsection{Acceptance criterion}
\label{sec:acceptance-criteria}

The convergence threshold $f^*$ is the smallest partition fraction at which the $1\sigma$ band of $\log_{10}R$ lies inside $\pm 0.05$ dex in every bin of the range considered, and continues to do so at every larger fraction measured. At $f^*$, therefore, at least 68 per cent of selections lie within $0.05$ dex in each bin.
The bins are correlated but not perfectly (Section~\ref{sec:covariance}), so the fraction of selections within tolerance in all bins at once is lower than in the worst bin and can fall below 68 per cent. A selection returning $\hat\xi \leq 0$ in a bin lies outside the band and counts as a failure.

The tolerance, $0.05$ dex, is a factor of $10^{0.05} = 1.12$, or 12 per cent. We adopt it to define convergence; calibration against data may need a tighter tolerance, as discussed in item~(ii) of Section~\ref{sec:conclusions}.

A selection's corrected pair count is a sum of the auto-pair counts of its partitions and the cross-pair counts of its partition pairs, so we count pairs once for every pair of the $k$ partitions and obtain any selection by summation.
This gives $f^\ast$ from 10,500 selections at every integer $m$ from 2 to $k-1$; the first 105 are those used elsewhere in the paper. Depth matters: requiring acceptance at every larger fraction lets one noisy failure at large $m$ push $f^\ast$ up, so 105 selections give $f^\ast$ typically 15, and up to 40, per cent higher. At 10,500 the bootstrap 68 per cent interval on $f^\ast$ is within $\pm3$ per cent, and halves and quarters of the ensemble agree to within 3.4 and 7 per cent. We quote $f^\ast$ to two significant figures.

The criterion cannot be met where selections often contain no pair in a bin, since the weights cannot reweight an empty count. This happens at the smallest separations and smallest $m$ in the sparser samples (Appendix~\ref{app:positivity-censoring}), and is why the sparsest sample, $\bar n = 10^{-4}\,h^3\,\mathrm{Mpc}^{-3}$, meets the criterion over the full range only above $f = 96$ per cent, where \ac{SaUCE} saves almost nothing.

\subsection{Redshift-space multipoles}
\label{sec:rsd-metrics}

In redshift space the reference is the monopole $\xi_{0,\rm ref}(s)$ and quadrupole $\xi_{2,\rm ref}(s)$ of the merged $k$-partition catalogue, from equation~\eqref{eq:multipoles}. We bin pairs in $s$ on the grid of Section~\ref{sec:bins} and in $\mu$ in $24$ linear bins over $0 \leq \mu \leq 1$; the periodic box with a fixed line of sight gives $\xi(s,\mu) = \xi(s,-\mu)$, so negative $\mu$ adds nothing.

The quadrupole changes sign near the one-halo to two-halo transition, so its log-ratio is undefined there. Both multipoles are therefore assessed on the residual normalised by the monopole,
\begin{equation}
e_\ell(s) = \frac{\xi_\ell(s) - \xi_{\ell,\rm ref}(s)}{\xi_{0,\rm ref}(s)}, \qquad \ell = 0,\,2,
\label{eq:rsd-metric}
\end{equation}
which is defined where $\xi_{2,\rm ref}$ crosses zero. We apply the criterion of Section~\ref{sec:acceptance-criteria} to the 16th--84th percentile band of $\log_{10}(1 + e_\ell)$, in dex. For the monopole, $1 + e_0 = \xi_0/\xi_{0,\rm ref}$, so this is the criterion used for $\xi(r)$, with $\xi_0$ in place of $\xi$.

\section{Convergence and accuracy of SaUCE}\label{sec:results}

Section~\ref{sec:convergence} shows how the estimated 2PCF depends on the partition fraction. Section~\ref{sec:sample-selection} extends these results to different galaxy samples and redshifts. Section~\ref{sec:model-robustness} shows that the scatter is the same for two \texttt{GALFORM} variants, and Section~\ref{sec:box-size-dependence} whether a partition fraction transfers between boxes. Section~\ref{sec:variance-recipe} gives a practical recipe for predicting the estimator's variance, and Section~\ref{sec:covariance} characterises its bin-to-bin covariance. Section~\ref{sec:rsd-results} tests the correction in redshift space, and asks what each multipole costs. Partition fractions quoted here are specific to the P-Millennium unless stated otherwise.

\subsection{Convergence with partition fraction}
\label{sec:convergence}

Fig.~\ref{fig:method-improvement} shows the effect of the correction for ${\bar{n} = 10^{-2}\,h^3\,\mathrm{Mpc}^{-3}}$ at $z = 0$. The naive estimate overstates $\xi(r)$ on small scales by the factor $k/m$ predicted in Section~\ref{sec:bias},
while on large scales, where almost all pairs are two-halo pairs, the overstatement vanishes.

\begin{figure*}
	\centering
	\offinterlineskip
	\hbox{\includegraphics[width=0.306752\textwidth]{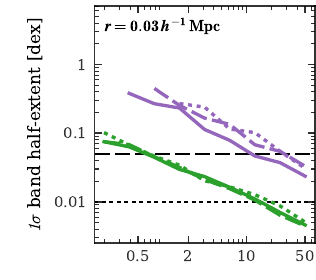}\includegraphics[width=0.225730\textwidth]{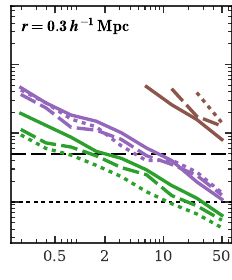}\includegraphics[width=0.225730\textwidth]{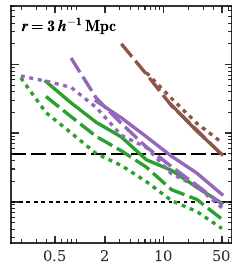}\includegraphics[width=0.241788\textwidth]{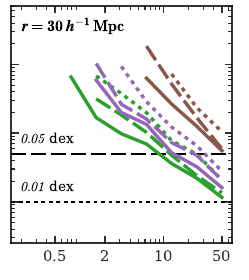}}
	\hbox{\includegraphics[width=\textwidth]{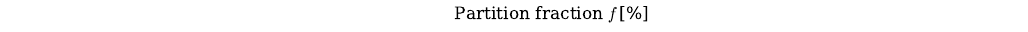}}
	\vspace{2pt}
	\centerline{\includegraphics{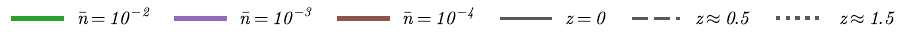}}

	\caption{The scatter of \ac{SaUCE} against the partition fraction $f$, one radial bin per panel (P-Millennium, \texttt{L16}). The panels are the bins centred on $r = 0.03$, $0.3$, $3$ and $30\,h^{-1}\,\mathrm{Mpc}$, and the nine curves in each are three number-density samples at three redshifts (legend); the $r = 0.03\,h^{-1}\,\mathrm{Mpc}$ panel omits the sparsest sample, whose $\hat\xi$ is too often $\leq 0$ there. The ordinate is the half-extent of the $1\sigma$ band of $\log_{10}R$, so a curve crosses the dashed $0.05$ dex line where that bin meets the criterion of Section~\ref{sec:acceptance-criteria}; the dotted line marks $0.01$ dex. A selection returning $\hat\xi \leq 0$ counts as a failure (Section~\ref{sec:acceptance-criteria}). $m = 1$ is excluded, since $w_{\rm cross}$ is undefined for a single partition.}

	\label{fig:convergence-regimes}
\end{figure*}

The estimator is unbiased, and what improves with $f$ is the scatter. The mean ratio $\langle R\rangle$ of equation~\eqref{eq:rho} stays within a few per cent of unity at every $f$ and separation, with no trend in $f$, consistent with the noise on the mean (Appendix~\ref{app:mean-vs-median}). The scatter falls steeply and smoothly with $f$ (Fig.~\ref{fig:convergence-regimes}), and for $\bar n = 10^{-2}\,h^3\,\mathrm{Mpc}^{-3}$ is smaller at small separations than at large ones.

Which scales set the threshold depends on the sample (Table~\ref{tab:convergence}). Usually it is the two-halo range. For $\bar n = 10^{-3}\,h^3\,\mathrm{Mpc}^{-3}$ it is the smallest scales, $r < 0.3\,h^{-1}\,\mathrm{Mpc}$, where that sample has few pairs per bin, though at every redshift only narrowly ahead of the two-halo range: $15$, $21$ and $29$ per cent against $14$, $20$ and $28$. The one-halo range is never limiting.

\subsection{Generalisation across samples and redshifts}
\label{sec:sample-selection}

\begin{figure}
	\centering
	\includegraphics[width=\columnwidth]{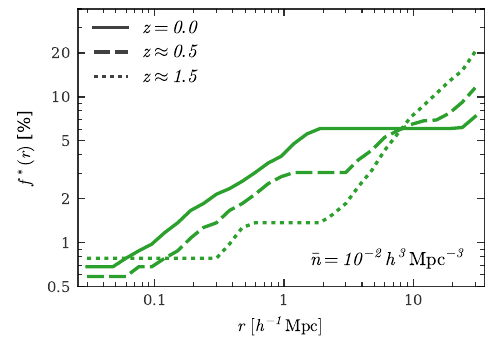}%
	\caption{The partition fraction needed as a function of the largest separation used, for $\bar n = 10^{-2}\,h^3\,\mathrm{Mpc}^{-3}$ on the P-Millennium. The ordinate is the largest per-bin threshold over all bins from $0.03\,h^{-1}\,\mathrm{Mpc}$ out to $r$, not the threshold of the bin at $r$. $\bar n = 10^{-3}\,h^3\,\mathrm{Mpc}^{-3}$ is not shown, since its smallest scales set the full-range requirement and its curves would be flat (Table~\ref{tab:convergence}).}
	\label{fig:fstar-vs-r}
\end{figure}

\begin{table}
\centering
\caption{The convergence threshold $f^*$ for the P-Millennium, from $10{,}500$ selections at every $m$: the smallest partition fraction at which the $1\sigma$ band of $\log_{10}R$ lies inside $\pm 0.05$ dex in every radial bin of the stated range, and at every larger fraction measured (Section~\ref{sec:acceptance-criteria}). The full range is ${0.03 \leq r \leq 30\,h^{-1}\,\mathrm{Mpc}}$. The one-halo ($0.3$--$1$), transition ($1$--$5$) and two-halo ($5$--$30\,h^{-1}\,\mathrm{Mpc}$) columns show which scales set it; the full range also contains the smallest scales, $r < 0.3\,h^{-1}\,\mathrm{Mpc}$, which set it where it exceeds all three columns. $m$ is the partition count, $f^* = m/k$ with $k = 1024$, and $\bar n$ is in $h^3\,\mathrm{Mpc}^{-3}$. The sparsest sample, $\bar n = 10^{-4}\,h^3\,\mathrm{Mpc}^{-3}$, meets the criterion over the full range only above $f = 96$ per cent.}
\label{tab:convergence}
{\setlength{\tabcolsep}{4pt}
\begin{tabular}{ll cc ccc}
\hline
 & & \multicolumn{2}{c}{full range} & \multicolumn{3}{c}{sub-range $f^*$ [per cent]} \\
\cmidrule(lr){3-4}\cmidrule(lr){5-7}
$z$ & $\bar n$ & $m$ & $f^*$ [per cent] & one-halo & transition & two-halo \\
\hline
$0$ & $10^{-2}$ & $75$  & $7.3$ & $3.9$ & $6.1$  & $7.3$ \\
$0$ & $10^{-3}$ & $157$ & $15$ & $9.6$ & $13$  & $14$ \\
$0.5$ & $10^{-2}$ & $118$ & $12$ & $2.8$ & $4.3$  & $12$ \\
$0.5$ & $10^{-3}$ & $212$ & $21$ & $8.1$ & $9.3$  & $20$ \\
$1.5$ & $10^{-2}$ & $209$ & $20$ & $1.4$ & $3.1$  & $20$ \\
$1.5$ & $10^{-3}$ & $299$ & $29$ & $7.3$ & $7.7$  & $28$ \\
\hline
\end{tabular}}
\end{table}

\begin{figure*}
	\centering
	\offinterlineskip
	\hbox{\includegraphics[width=0.297263\textwidth]{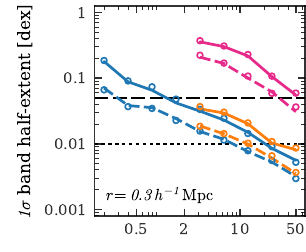}\includegraphics[width=0.225000\textwidth]{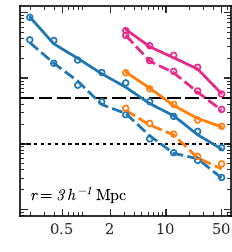}\includegraphics[width=0.236679\textwidth]{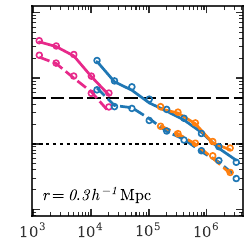}\includegraphics[width=0.241058\textwidth]{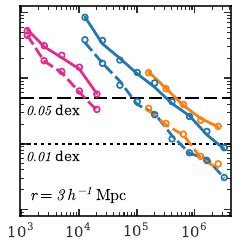}}
	\hbox{\includegraphics[width=\textwidth]{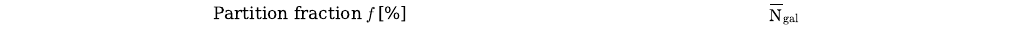}}
	\vspace{2pt}
	\centerline{\includegraphics{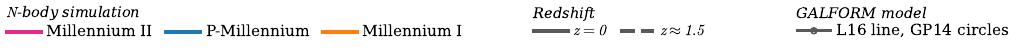}}

	\caption{The scatter of \ac{SaUCE} for two galaxy formation models in three boxes, for $\bar n = 0.0408\,h^3\,\mathrm{Mpc}^{-3}$. Every panel has the ordinate of Fig.~\ref{fig:convergence-regimes}, the half-extent of the $1\sigma$ band at a single radial bin, with the $0.05$ dex tolerance dashed and $0.01$ dex dotted. The abscissa is the partition fraction $f$ in the left pair of panels and the galaxy count in the selected partitions, $\overline{\mathrm{N}}_{\rm gal}$, in the right pair. Within each pair the panels are $r = 0.3$ and $3\,h^{-1}\,\mathrm{Mpc}$. Colour is the simulation (P-Millennium $L = 542.2$, Millennium~I $L = 500$, Millennium~II $L = 100\,h^{-1}\,\mathrm{Mpc}$), line style the redshift (solid $z = 0$, dashed $z \approx 1.5$), and the model is \texttt{L16} for the lines and \texttt{GP14} for the circles on them. Against $\overline{\mathrm{N}}_{\rm gal}$ the boxes fall on one relation at $r = 0.3\,h^{-1}\,\mathrm{Mpc}$ but not at $3\,h^{-1}\,\mathrm{Mpc}$ (Section~\ref{sec:box-size-dependence}).}

	\label{fig:model-robustness}
\end{figure*}

Fig.~\ref{fig:convergence-regimes} shows how the scatter depends on the sample and redshift: within a panel, curves of one line style differ only in sample and curves of one colour only in redshift, and comparing panels isolates the scale.

The required fraction rises as the number density falls, because a sparser sample has fewer pairs in each bin. Over the full range and on two-halo scales it also rises with redshift at fixed $\bar n$. The number of galaxies is then the same, but the two-halo clustering amplitude is lower (near $10\,h^{-1}\,\mathrm{Mpc}$, $\xi_{\rm ref}$ is about a third lower at $z \approx 1.5$ than at $z = 0$), so the same scatter in the pair count is a larger fraction of $\xi$. On one-halo and transition scales the threshold instead falls with redshift, for both samples in Table~\ref{tab:convergence}: for $\bar n = 10^{-2}\,h^3\,\mathrm{Mpc}^{-3}$ it falls from $3.9$ to $1.4$ per cent on one-halo scales and from $6.1$ to $3.1$ per cent on transition scales between $z = 0$ and $z \approx 1.5$. We return to the cause in Section~\ref{sec:box-size-dependence}.

Table~\ref{tab:convergence} gives the fraction needed over the full range. A user who needs only smaller separations can run much less: Fig.~\ref{fig:fstar-vs-r} gives, against $r$, the fraction needed for every bin from $0.03\,h^{-1}\,\mathrm{Mpc}$ out to $r$ to meet the criterion.

\subsection{Galaxy formation models}
\label{sec:model-robustness}

The bias of \ac{SaUCE} does not depend on the galaxy formation model (Section~\ref{sec:sauce-derivation}), but its scatter could. We compare the \texttt{L16} and \texttt{GP14} models, which have different feedback prescriptions (Appendix~\ref{app:galaxy-formation-models}), in all three simulations, with galaxies selected to $\bar{n} = 0.0408\,h^3\,\mathrm{Mpc}^{-3}$ (Fig.~\ref{fig:model-robustness}). The \texttt{GP14} circles lie on the \texttt{L16} lines throughout: the bands of the two models differ by no more than the uncertainty of a band measured from $105$ selections (bin by bin, the \texttt{L16}/\texttt{GP14} band ratio has median $0.94$--$1.00$ in each box and redshift), and their thresholds agree to within one step of the power-of-two grid in $m$, a factor of two in $f$. The scatter is therefore the same for these two \texttt{GALFORM} variants, although their full-box correlation functions differ by up to $8$ per cent in the P-Millennium at $z = 0$, and by less than $3$ per cent at $z \approx 1.5$ (Appendix~\ref{app:galaxy-formation-models}). Matched in number density, however, their \acsp{HOD} are similar: for this sample in Millennium~I, at both redshifts, the halo masses at which $\langle N_{\rm cen}\rangle = 1/2$ and $\langle N_{\rm sat}\rangle = 1$ agree to within $0.02$ dex, and $\langle N_{\rm sat}\rangle$ above the $\langle N_{\rm sat}\rangle = 1$ mass to within about $10$ per cent.

The test therefore shows that the scatter is insensitive to differences in galaxy formation physics at fixed occupation; models whose HODs differ more, for example in satellite fraction, could have a different
scatter.

\subsection{Box size}
\label{sec:box-size-dependence}

We now ask whether a partition fraction measured in one box transfers to another. Millennium~I and Millennium~II both have $k = 64$ partitions, and we test powers of two in $m$ over $0.03$--$7.5\,h^{-1}\,\mathrm{Mpc}$, the range all three boxes share, on the $0.1$ dex grid of Section~\ref{sec:bins}. Millennium~I meets the criterion at a similar fraction to the P-Millennium, within one step in $m$, a factor of two in $f$. Millennium~II, whose volume is $125$ times smaller, never meets it, even at $f = 50$ per cent.

The cause is the small number of galaxies per partition. Each partition holds a fraction $1/k$ of the galaxies while spanning the whole volume (Section~\ref{sec:merger-tree-partitioning}), and all boxes are selected to the same $\bar n$, so each Millennium~II partition holds $125$ times fewer galaxies than a Millennium~I partition. Millennium~II's scatter keeps falling as $f$ increases, as expected for sampling noise; the corrected pair counts are unbiased in any box (Section~\ref{sec:sauce-derivation}).

The right-hand panels of Fig.~\ref{fig:model-robustness} plot the scatter against the mean number of galaxies in the selected partitions, $\overline{\mathrm{N}}_{\rm gal} = \bar n\, f\, L_{\rm box}^3$, instead of $f$. The three boxes then fall on one relation at $r = 0.3\,h^{-1}\,\mathrm{Mpc}$: there the scatter is set by whether a halo enters the selection with its satellites, which depends only on how many galaxies are drawn. At $r = 3\,h^{-1}\,\mathrm{Mpc}$ they do not, because the pairs come from distinct haloes and their number depends on the density of the selection as well as its size. The number of pairs in a shell of volume $V_{\rm s}$ is $\overline{\mathrm{N}}_{\rm gal}^2 V_{\rm s}/(2L_{\rm box}^3)$, so $\overline{\mathrm{N}}_{\rm gal}^2/L_{\rm box}^3$ measures the pair count per unit shell volume; plotted against it (not shown), the boxes fall on one relation at $r = 3\,h^{-1}\,\mathrm{Mpc}$, their median offsets from a single power law spanning a factor of $1.3$, compared with $2.3$ for $\overline{\mathrm{N}}_{\rm gal}$. A larger box therefore needs a smaller partition fraction for the same scatter: $f$ transfers between boxes through the galaxy count on small scales and through the pair count per unit shell volume on large scales.

In every panel of Fig.~\ref{fig:model-robustness} the $z \approx 1.5$ band lies below the $z = 0$ band, and neither variable removes this: at these separations one-halo pairs, which enter a selection in bursts, are a smaller share of the total at higher redshift, as in the one-halo and transition columns of Table~\ref{tab:convergence}.

\subsection{A recipe for the estimator's scatter}
\label{sec:variance-recipe}

\begin{figure*}[b]
	\centering
	\includegraphics[width=0.488321\textwidth]{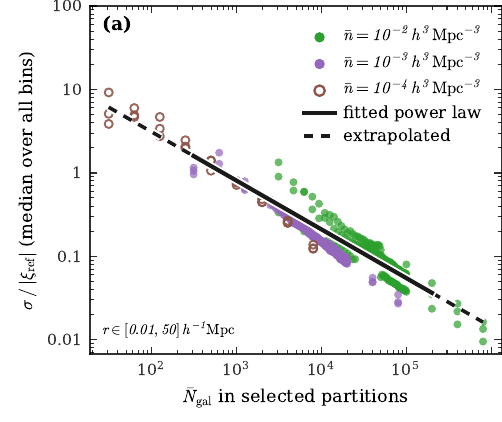}%
	\includegraphics[width=0.511679\textwidth]{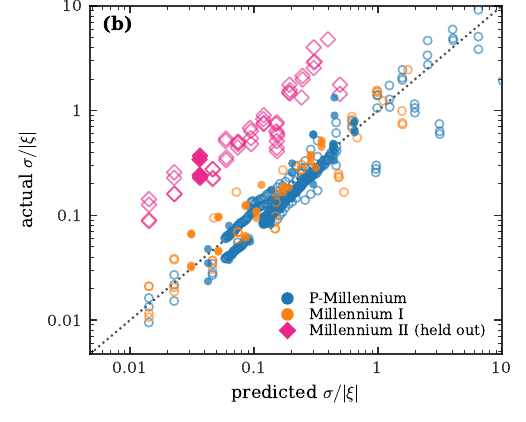}
	\caption{The fractional scatter $\sigma/|\xi_{\rm ref}|$ of the \ac{SaUCE} estimator and the recipe for predicting it (Section~\ref{sec:variance-recipe}). \textit{Left}: the median $\sigma/|\xi_{\rm ref}|$ over the radial bins, $0.01 \leq r < 50\,h^{-1}\,\mathrm{Mpc}$ in $0.148$ dex bins, against the galaxy count in the selected partitions, $\overline{\mathrm{N}}_{\rm gal}$, for every density, redshift and $f$ of the P-Millennium. The solid line is equation~\eqref{eq:variance-recipe-ngal} fitted to the $\bar n = 10^{-2}$ and $10^{-3}\,h^3\,\mathrm{Mpc}^{-3}$ samples (filled) at $f \leq 12.5$ per cent; the dashed line extends it. The sparsest sample, $\bar n = 10^{-4}\,h^3\,\mathrm{Mpc}^{-3}$ (open), is not fitted. \textit{Right}: measured against predicted scatter for equation~\eqref{eq:variance-recipe-general}, fitted on the P-Millennium and Millennium~I. Each filled circle is predicted by the fit to the other simulation; open circles are measurements outside the fitted range, shown against the fit to both; Millennium~II (diamonds) is not fitted but predicted by the fit to both; the four filled diamonds (two models, two redshifts; the models nearly coincide) pass the cuts of Table~\ref{tab:variance-recipe} and give the comparison of Section~\ref{sec:variance-recipe}. The dotted line is $1{:}1$. Both axes are capped at $\sigma/|\xi_{\rm ref}| = 10$, which excludes one of the $649$ points. Colour is number density on the left and simulation on the right.}
		\label{fig:variance-collapse}
\end{figure*}

The closed-form variance of Appendix~\ref{app:closed-form-variance} confirms that the measured scatter is what the random choice of partitions predicts, but a user cannot evaluate it: it needs the full-box pair counts that \ac{SaUCE} exists to avoid. We therefore fit the scatter empirically. Pair separation alone does not predict it: at fixed $r$, the fractional scatter $\sigma/|\xi_{\rm ref}|$ differs between number densities (Fig.~\ref{fig:variance-validation}a, at $z = 0$) and between redshifts (Fig.~\ref{fig:convergence-regimes}, whose band width follows from this variance; Section~\ref{sec:accuracy-metrics}), because $r$ says nothing about how many pairs contribute to a bin. Here $\sigma$ is the standard deviation of $\hat{\xi}$ over selections at fixed $f$, measured bin by bin and reduced to one number per fitted row by the median over the bins of the range in question.

We fit two forms. The first is a power law in the galaxy count $\overline{\mathrm{N}}_{\rm gal}$ of Section~\ref{sec:box-size-dependence},

\begin{equation}
\frac{\sigma}{|\xi_{\rm ref}|} = A_N\, \overline{\mathrm{N}}_{\rm gal}^{\,p_N},
\label{eq:variance-recipe-ngal}
\end{equation}
and the second separates $f$ and $\bar n$,

\begin{equation}\label{eq:variance-recipe-general}
\frac{\sigma}{|\xi_{\rm ref}|} = A\, f^{\,B}\, \bar n^{\,C}.
\end{equation}

\begin{table}
\centering
\caption{Coefficients for the all-scale recipes, fitted to the median fractional scatter over $0.01 \leq r < 50\,h^{-1}\,\mathrm{Mpc}$. The rms is the residual scatter in $\log_{10}(\sigma/|\xi_{\rm ref}|)$. In equation~\eqref{eq:variance-recipe-general}, use $f = m/k$ as a fraction and $\bar n$ in $h^3\,\mathrm{Mpc}^{-3}$. The P-Millennium-only fit uses the $\bar n = 10^{-2}$ and $10^{-3}\,h^3\,\mathrm{Mpc}^{-3}$ samples and ${m = 2,4,\ldots,128}$; the two-simulation fits use the $\bar n = 10^{-2}$--$10^{-4}$ samples with power-of-two $m \geq 2$, $f \leq 0.125$ and selected galaxy counts above $1000$.}
\label{tab:variance-recipe}
{\setlength{\tabcolsep}{4pt}
\begin{tabular}{lrrr}
\hline
\multicolumn{4}{l}{Equation~\eqref{eq:variance-recipe-ngal}} \\
Calibration & $A_N$ & $p_N$ & rms [dex] \\
\hline
P-Millennium & $45.6$ & $-0.584$ & $0.176$ \\
P-Millennium + Millennium~I & $38.0$ & $-0.570$ & $0.159$ \\
\hline
\multicolumn{4}{l}{Equation~\eqref{eq:variance-recipe-general}, P-Millennium + Millennium~I} \\
$A$ & $B$ & $C$ & rms [dex] \\
\hline
$8.32\times10^{-4}$ & $-0.677$ & $-0.514$ & $0.136$ \\
\hline
\end{tabular}}
\end{table}

We fit $A_N$, $p_N$, $A$, $B$ and $C$. Within one box $\overline{\mathrm{N}}_{\rm gal} \propto \bar n f$, so equation~\eqref{eq:variance-recipe-ngal} is equation~\eqref{eq:variance-recipe-general} with $B = C$; what it adds is $L_{\rm box}^3$, which lets it cross box sizes. Equation~\eqref{eq:variance-recipe-general} tests whether $B = C$, and gives the dependence on $f$ at fixed $\bar n$. We fit both by least squares in $\log_{10}$, over all separations and separately on one-halo ($r < 1$), transition ($1 \leq r < 5$) and two-halo ($5 \leq r < 50\,h^{-1}\,\mathrm{Mpc}$) scales, ranges wider than those of Section~\ref{sec:accuracy-metrics}; Table~\ref{tab:variance-recipe} gives the all-scale coefficients and selection cuts.
The fits use powers of two in $m$ on the coarser grids of Section~\ref{sec:bins},
each over at least 105 selections.

Intermediate partition counts were run only inside brackets around the $0.05$ dex crossings of a preliminary, $105$-selection estimate of the thresholds; they sample $m$ unevenly, so they are not used in fitting. The recipe predicts the typical scatter, the median over bins, whereas the criterion of Section~\ref{sec:acceptance-criteria} is set by the worst bin, and equation~\eqref{eq:variance-recipe-general} has no redshift term. A single bin meets a tolerance of $t$ dex when $\sigma/|\xi_{\rm ref}| \lesssim t \ln 10$, or $0.115$ for $t = 0.05$ dex, but at the thresholds of Table~\ref{tab:convergence} equation~\eqref{eq:variance-recipe-general} gives (extrapolating above $f = 0.125$) a median $\sigma/|\xi_{\rm ref}|$ of only $0.026$--$0.103$, so inverting it at $0.115$ would underestimate $f^*$ by a factor of up to about nine. No single target fixes this, because the factor ranges from $1.2$ to $9$ across Table~\ref{tab:convergence}. The partition fraction is therefore taken from Table~\ref{tab:convergence} and Fig.~\ref{fig:fstar-vs-r} where they apply, and otherwise from repeated selections.

The left panel of Fig.~\ref{fig:variance-collapse} shows that within one box $\overline{\mathrm{N}}_{\rm gal}$ predicts the scatter: pooling every density, redshift and $f$ of the P-Millennium, the $\bar n = 10^{-2}$ and $10^{-3}\,h^3\,\mathrm{Mpc}^{-3}$ samples fall on one power law with $p_N = -0.58$ and an rms scatter about the fit of $0.18$ dex, which is the accuracy to which the recipe predicts $\sigma$. At fixed $\bar n$ and box size $\sigma \propto f^{\,p_N}$, so this is an rms of $0.18/0.58 \approx 0.3$ dex, a factor of two, in the $f$ that gives a given typical scatter. Fitted range by range, the exponent is close to the value $-1/2$ expected for independent counts only on one-halo scales, where the scatter is set by how many galaxies are drawn; on larger scales the scatter falls faster with $\overline{\mathrm{N}}_{\rm gal}$. The exponent depends on the range of $f$ fitted. At larger $f$ the measurements lie below the fit, since the scatter must reach zero at $f = 1$, so the recipe is conservative at large $f$.

For the P-Millennium-only fit, the sparsest sample is not fitted. Its points with $\sigma < |\xi_{\rm ref}|$ lie on the extrapolated relation; points with $\sigma > |\xi_{\rm ref}|$, in any sample, lie above it. The recipe therefore applies where the scatter is smaller than the signal. Where the predicted $\sigma$ approaches $|\xi|$, $\sigma$ must come from repeated selections.

Pooled over all scales, and on one-halo scales, $B$ and $C$ differ by less than 0.2. On two-halo scales $B < C$ in each simulation separately ($B - C = -0.63$ and $-0.68$): spreading a fixed number of galaxies over more partitions lowers the two-halo scatter, since two-halo pairs lie mostly between partitions, so there $f$ and $\bar{n}$ do not act only through $\overline{\mathrm{N}}_{\rm gal}$.

The right panel of Fig.~\ref{fig:variance-collapse} tests whether the recipe transfers between simulations. Fitted over all scales on the P-Millennium and Millennium~I together, equation~\eqref{eq:variance-recipe-general} gives $B = -0.68$ and $C = -0.51$. Predicting each simulation from the fit to the other raises the rms from $0.14$ to $0.16$ dex, across a change of cosmology and partition count; divided by $|B|$, these are factors of $1.6$ and $1.7$ in $f$.

The data cannot constrain a box-size term: the P-Millennium and Millennium~I have similar box sizes, and only four Millennium~II measurements pass the cuts of Table~\ref{tab:variance-recipe}, all at $f = 0.125$ and $\bar n = 10^{-2}\,h^3\,\mathrm{Mpc}^{-3}$ (two models at two redshifts), a single location in $(f, \bar n)$. The two-simulation fit underestimates Millennium~II's $\sigma$ (filled diamonds in Fig.~\ref{fig:variance-collapse}; its medians are taken over 0.01--10\,$h^{-1}$\,Mpc, the edge of its grid) by about a factor of eight, overstating the precision. Equation~\eqref{eq:variance-recipe-ngal}, refitted on the same two simulations, contains the box size through $\overline{\mathrm{N}}_{\rm gal}$, and overestimates it by about a factor of $2.3$. We therefore recommend equation~\eqref{eq:variance-recipe-general} for boxes of $500$--$542\,h^{-1}\,\mathrm{Mpc}$, and equation~\eqref{eq:variance-recipe-ngal} for other box sizes.

\subsection{Bin-to-bin covariance}
\label{sec:covariance}

\begin{figure*}
	\centering
	\begin{minipage}[b]{0.477372\textwidth}
		\offinterlineskip
		\hbox{\includegraphics[width=0.061162\linewidth]{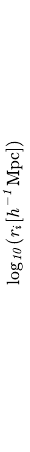}\vbox{\offinterlineskip\hbox{\includegraphics[width=0.474006\linewidth]{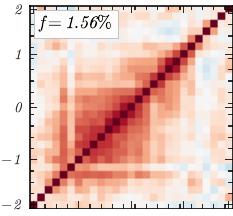}\includegraphics[width=0.464832\linewidth]{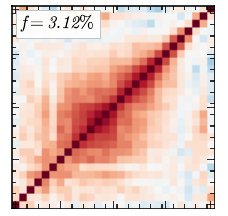}}\hbox{\includegraphics[width=0.474006\linewidth]{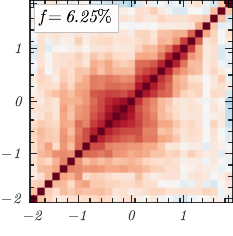}\includegraphics[width=0.464832\linewidth]{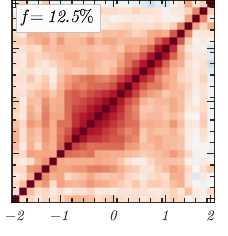}}}}
		\hbox{\includegraphics[width=\linewidth]{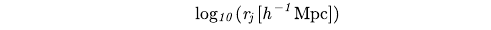}}
	\end{minipage}%
	\includegraphics[width=0.064234\textwidth]{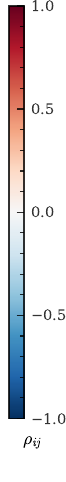}%
	\includegraphics[width=0.458394\textwidth]{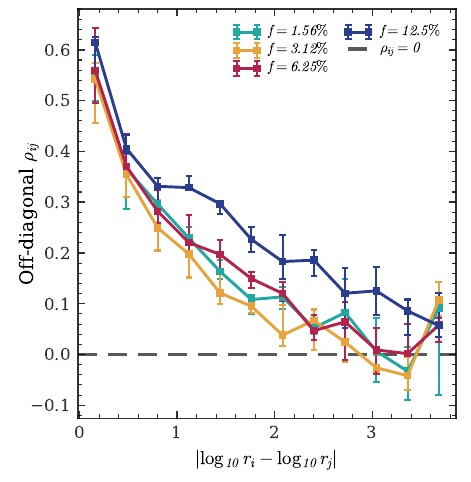}

	\caption{Bin-to-bin correlation matrix $\rho_{ij}$ of the \ac{SaUCE} estimator across selections ($z \approx 1.5$, $\bar n = 10^{-2}\,h^3\,\mathrm{Mpc}^{-3}$, $0.148$ dex bins), for $f = 1.56$, $3.12$, $6.25$ and $12.5$ per cent, from $100$ selections each, the same number at every $f$ so that the panels are directly comparable. The final panel shows the median off-diagonal correlation against bin separation $|\log_{10}r_i - \log_{10}r_j|$. Error bars are the 16th--84th percentiles of $800$ bootstrap resamplings of the bin pairs contributing to each point, the uncertainty on the median; bin pairs sharing a radial bin are not independent, so the bars are a lower bound. Separation bins holding fewer than four bin pairs are omitted.}

	\label{fig:corr-matrix}
\end{figure*}

\begin{figure*}[t]
	\centering
	\offinterlineskip
	\hbox{\includegraphics[width=0.359124\textwidth]{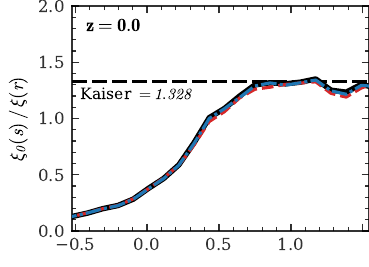}\includegraphics[width=0.310949\textwidth]{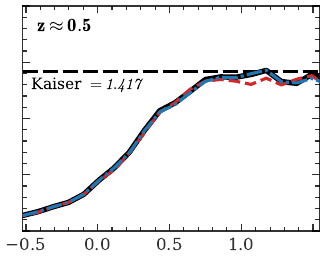}\includegraphics[width=0.329927\textwidth]{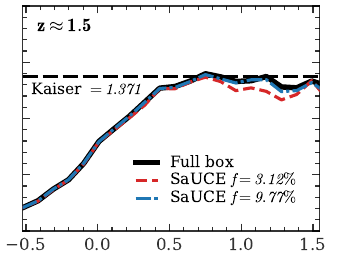}}
	\hbox{\includegraphics[width=\textwidth]{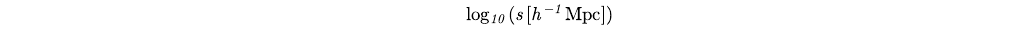}}

	\caption{Ratio of the redshift-space monopole to the real-space correlation function, $\xi_0(s)/\xi(r)$, for $\bar n = 10^{-2}\,h^3\,\mathrm{Mpc}^{-3}$ at the three redshifts annotated. Numerator and denominator come from the same selection. The reference is thick black, beneath the \ac{SaUCE} estimates at $f = 3.12$ per cent (red dashed) and $9.77$ per cent (blue dash-dotted). The horizontal dashed line, annotated in each panel, is the linear-theory Kaiser boost $1 + 2\beta_{\rm Kaiser}/3 + \beta_{\rm Kaiser}^2/5$, with $\beta_{\rm Kaiser} = (\mathrm{d}\ln D/\mathrm{d}\ln a)/b$, $D$ the linear growth factor and the galaxy bias $b = (\xi_{\rm gal}/\xi_{\rm DM})^{1/2}$ measured over $10$--$30\,h^{-1}\,\mathrm{Mpc}$ with $\xi_{\rm DM}$ from linear theory.}

	\label{fig:rsd-kaiser-bias}
\end{figure*}

The scatter characterised above is not independent between separation bins, and Fig.~\ref{fig:corr-matrix} shows the bin-to-bin correlation matrix $\rho_{ij}$ of the \ac{SaUCE} estimator, measured across selections at $z \approx 1.5$ for the $\bar n = 10^{-2}\,h^3\,\mathrm{Mpc}^{-3}$ sample. A block of strongly correlated bins, with $\rho_{ij} \gtrsim 0.5$, extends across the one-halo to two-halo scales ($r \sim 0.1$--$10\,h^{-1}\,\mathrm{Mpc}$) at every partition fraction shown. This structure is a consequence of the partition sampling: adding or removing a partition changes the galaxy content coherently across a wide range of separations, so a selection that scatters high in one bin tends to scatter high in its neighbours. The correlation is therefore a property of which haloes are selected rather than of the pair counting within a fixed catalogue. Each $\rho_{ij}$ carries noise of ${\approx}0.1$ from $100$ selections, so the isolated high and low cells in the largest-separation corners, where pair counts per partition are low, are noise. The final panel quantifies how that coherence decays, plotting the median off-diagonal correlation against bin separation $|\log_{10} r_i - \log_{10} r_j|$. The three sparser fractions track one another closely; $f = 12.5$ per cent decays more slowly, staying some $0.1$ higher at every separation out to $2$ dex.

How many independent quantities the bins carry is measured by the effective number of bins, $N_{\rm eff} = (\sum_a\lambda_a)^2/\sum_a\lambda_a^2$, where $\lambda_a$ is the $a$-th eigenvalue of $\rho_{ij}$ and the sums run over all of them. $N_{\rm eff}$ equals the number of bins when the bins are uncorrelated and falls to one when they are perfectly correlated. We compute $N_{\rm eff}$ from the correlation matrices of Fig.~\ref{fig:corr-matrix}, once for all $27$ bins and once for each clustering regime of Section~\ref{sec:accuracy-metrics}, restricting $\rho_{ij}$ to that regime's bins; the ranges quoted span the four fractions shown. Across all $27$ bins $N_{\rm eff} = 6.0$--$7.8$. The one-halo and transition regimes, four bins each on this grid, give $1.3$--$1.5$ and $1.6$--$2.0$, and the two-halo regime, six bins, gives $2.6$--$3.4$.

Subdividing the radial range more finely therefore adds little independent information about the partition-selection scatter: the additional bins are almost perfectly correlated with their neighbours, raising the nominal bin count without raising $N_{\rm eff}$.

A likelihood must therefore use the full covariance: a diagonal approximation would count correlated fluctuations as independent evidence and give artificially tight constraints. The correlation structure changes only slowly with $f$ and may be reused across the range used for calibration, but should be estimated at the largest $f$ in use, since denser selections retain correlation to larger separations.

The matrix of Fig.~\ref{fig:corr-matrix} is the covariance of the estimator across selections: it captures the uncertainty introduced by choosing $m$ of $k$ partitions in a fixed simulation, and is measured cheaply by repeating the selection. It does not contain the cosmic variance of the parent $N$-body realisation, which is common to every selection, nor the covariance of the data; a full error budget must add both.

\subsection{Redshift-space distortions}
\label{sec:rsd-results}

Fig.~\ref{fig:rsd-kaiser-bias} shows the ratio of the redshift-space monopole to the real-space correlation function for $\bar n = 10^{-2}\,h^3\,\mathrm{Mpc}^{-3}$. \ac{SaUCE} tracks the reference at all three redshifts, including the finger-of-God suppression at small $s$, where the velocity dispersion within haloes stretches pairs along the line of sight, and the Kaiser boost on large scales \citep{Kaiser1987}.

At fixed $f$ the departure from the reference grows with redshift, for the reason given in Section~\ref{sec:sample-selection}.

\begin{figure}
	\centering
	\includegraphics{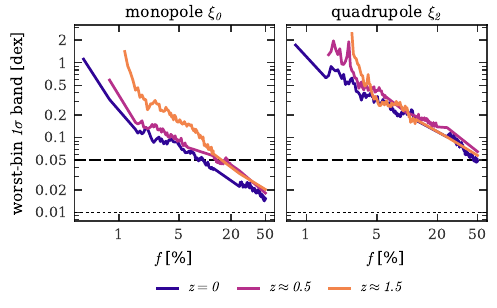}

	\caption{The scatter of the redshift-space multipoles against the partition fraction $f$ (P-Millennium, \texttt{L16}, $\bar n = 10^{-2}\,h^3\,\mathrm{Mpc}^{-3}$), at $z = 0$, $z \approx 0.5$ and $z \approx 1.5$ (colour). \textit{Left}: the monopole $\xi_0(s)$. \textit{Right}: the quadrupole $\xi_2(s)$, from the same selections. The ordinate is the half-extent of the $1\sigma$ band of $\log_{10}(1 + e_\ell)$ (Section~\ref{sec:rsd-metrics}) in the worst bin over $0.03 \leq s \leq 30\,h^{-1}\,\mathrm{Mpc}$, with $e_\ell$ from equation~\eqref{eq:rsd-metric}. A curve meets the criterion of Section~\ref{sec:acceptance-criteria} where it falls below the dashed $0.05$ dex line; the dotted line marks $0.01$ dex. At the smallest $f$ the band extends below $e_\ell = -1$, where $\log_{10}(1 + e_\ell)$ is undefined, so those points are not drawn.}

	\label{fig:quadrupole-convergence}
\end{figure}

Fig.~\ref{fig:quadrupole-convergence} shows the scatter of both multipoles, from the same selections as in real space. The multipoles were measured at powers of two in $m$ up to $512$ and at every integer $m$ in some of the intervals between them; where a threshold falls in an interval that was not filled, we quote the bracket. At a depth of $105$ selections, the monopole's full-range threshold is $9.5$ per cent at $z = 0$ and $18$ per cent at $z \approx 0.5$, compared with $8.2$ and $12$ per cent for $\xi(r)$. At $z \approx 1.5$ the bound is $12.5 < f^* \leq 25$ per cent, which contains the real-space threshold of $24$ per cent, so their ordering is unresolved. At $z \leq 0.5$ its requirement is set by the smallest separations, and at $z \approx 1.5$ by the two-halo range, as in real space. The quadrupole is far more expensive: over the full range it meets the criterion at $f^* \simeq 50$ per cent at $z = 0$, the top of the measured range, but not at the measured fractions up to $50$ per cent at the higher redshifts.

At every redshift the quadrupole's worst bins are the three smallest, ${s \leq 0.05\,h^{-1}\,\mathrm{Mpc}}$. Without them the threshold falls to $31$ per cent at $z = 0$, while $25 < f^* \leq 50$ per cent at the higher redshifts; intermediate partition counts within this bracket were not run. Range by range, the quadrupole needs more than the monopole everywhere, two to four times as much on one-halo scales.

The smallest scales are hardest because the quadrupole is a difference between the clustering along and across the line of sight. There that difference comes from the velocity dispersion within individual haloes, and \ac{SaUCE} keeps or discards each halo whole, so the massive, high-dispersion haloes that contribute most of it are either all in a selection or all out of it. We attribute to the same effect the cost of the monopole's smallest separations: at the same depth of $105$ selections, its bins at $s \leq 0.05\,h^{-1}\,\mathrm{Mpc}$ need $5.4$--$18$ per cent, against at most $1$ per cent for the same separations in real space (Fig.~\ref{fig:convergence-regimes}). A quadrupole analysis should therefore budget for the range of separations it uses, not for the monopole's fraction.

\FloatBarrier

\section{Conclusions}
\label{sec:conclusions}

SaUCE allows galaxy clustering to play a role in the calibration of galaxy formation models. The weights $w_{\rm auto}$ and $w_{\rm cross}$ (equation~7) depend only on $m$ and $k$, so the correction has no free parameters, and the corrected pair counts are exactly unbiased whatever model populates the haloes. We tested it with two \texttt{GALFORM} variants on three $N$-body simulations, at number densities from $10^{-4}$ to $4\times10^{-2}\,h^{3}\,{\rm Mpc}^{-3}$ and redshifts $z \le 1.5$. The results reduce to the following recipe.

(i) \textbf{Partition the trees and select the sample.} Assign whole merger trees to $k$ partitions, draw $m$ of them uniformly at random, and weight within- and between-partition pairs by $w_{\rm auto}$ and $w_{\rm cross}$ (Section~3.3); the weights apply unchanged to any binning of the correlation function, including $(s,\mu)$ and $w_{\rm p}(r_{\rm p})$. The galaxy formation model must process each tree independently, and whether a galaxy enters the sample should not depend on which partitions are drawn, except through a threshold set from the sparse catalogue (Section~3.5).

(ii) \textbf{Choose the tolerance.} We require at least 68 per cent of selections to lie within 0.05\,dex (12 per cent) of the full-box result in each bin. This is larger than the difference of up to 8 per cent in clustering between the two models used here (Appendix~B), so distinguishing models this close needs a tighter tolerance and a larger $f$.

(iii) \textbf{Read off the partition fraction where it has been measured.} In the P-Millennium, $f = 7.3$ per cent meets the 0.05\,dex tolerance over 0.03--30\,$h^{-1}$\,Mpc at $z = 0$ and $\bar{n} = 10^{-2}\,h^{3}\,{\rm Mpc}^{-3}$: fourteen times fewer trees to process. The fraction rises with redshift and as the galaxy number density falls, to 29 per cent at $z \approx 1.5$ and $\bar{n} = 10^{-3}\,h^{3}\,{\rm Mpc}^{-3}$ (Table~2); smaller separations alone need less (Fig.~4). For the sparsest sample, $10^{-4}\,h^{3}\,{\rm Mpc}^{-3}$, the full range needs 96 per cent, so SaUCE saves almost nothing unless the smallest separations are dropped.

(iv) \textbf{Otherwise, measure it.} Determine $f$ with repeated selections. Equations~(13) and~(14) (Section~5.5) show how the typical scatter changes with $f$, $\bar{n}$ and box size, and so which way to move from the nearest case in Table~2; between boxes, $f$ transfers through the galaxy count on small scales and through the pair count per unit shell volume on large scales (Section~5.4).

(v) \textbf{Use the full covariance.} The scatter is strongly correlated between bins (Section~5.6), so a likelihood must use the full covariance, measured at the largest $f$ in use, and must add to it the cosmic variance of the simulation and the covariance of the data.

(vi) \textbf{Redshift space.} The monopole costs up to 1.5 times as much as $\xi(r)$, the quadrupole far more, because its smallest separations are dominated by a few high-dispersion haloes; an analysis should budget for the separations it fits (Section~5.7).

We next plan to use SaUCE as the forward model in simulation-based inference, propagating the covariance measured here, and to apply it to other semi-analytic models, to halo occupation models, where the same weights hold, and to abundance matching, where matching on the sparse catalogue sets the threshold tested in Section~3.5. SaUCE is complementary to methods that reduce cost at other stages, such as the cosmological rescaling of merger trees used by \citet{Stiskalek2026}: rescaling reduces the cost of building trees for each cosmology, while SaUCE reduces how many are populated.

\section*{Acknowledgements}

OJH was supported by a UKRI PhD Studentship (grant number UKRI1738) and
Durham University. CMB, BL, SB and KW acknowledge support from STFC
(ST/X001075/1). SB is supported by a UKRI Future Leaders Fellowship
[MR/V023381/1 and UKRI2044]. BL is funded by the ERC Advanced Grant
UNCA (UKRI Frontiers Research Guarantee EP/Z533877/1).

This work used the DiRAC@Durham facility managed by the Institute for
Computational Cosmology on behalf of the STFC DiRAC HPC Facility
(\url{www.dirac.ac.uk}). The equipment was funded by BEIS capital
funding via STFC capital grants ST/K00042X/1, ST/P002293/1,
ST/R002371/1 and ST/S002502/1, Durham University and STFC operations grant ST/R000832/1.
DiRAC is part of the National e-Infrastructure.

OJH acknowledges the use of Large Language Models as a writing and
research aid in the preparation of this manuscript, including
literature searches, drafting assistance, and language editing; all
scientific content, analysis and conclusions are the work of the
authors.

\section*{Software}

The SaUCE package is written in Rust \citep{matsakis2014rust}, with a
Python interface \citep{vanrossum2009python}. This work also made use of Corrfunc \citep{sinha2019corrfunc_avx, sinha2020corrfunc} for pair counting, NumPy \citep{harris2020numpy}, SciPy \citep{virtanen2020scipy} and Matplotlib \citep{hunter2007matplotlib} for analysis and figure production, and NASA's Astrophysics Data System for bibliographic information.

The Rust/Python package implementing \ac{SaUCE} is released on PyPI as \texttt{pySaUCE} (\url{https://pypi.org/project/pySaUCE/}), with source hosted on GitHub at \url{https://github.com/OscarHickman/SaUCE} under the MIT licence.

\section*{Data Availability}
The \texttt{GALFORM} galaxy catalogues may be shared on reasonable request to the corresponding author.

\bibliographystyle{mnras}
\bibliography{correlation_stat}

\begin{thebibliography}{}
\makeatletter
\relax
\def\mn@urlcharsother{\let\do\@makeother \do\$\do\&\do\#\do\^\do\_\do\%\do\~}
\def\mn@doi{\begingroup\mn@urlcharsother \@ifnextchar [ {\mn@doi@}
  {\mn@doi@[]}}
\def\mn@doi@[#1]#2{\def\@tempa{#1}\ifx\@tempa\@empty \href
  {http://dx.doi.org/#2} {doi:#2}\else \href {http://dx.doi.org/#2} {#1}\fi
  \endgroup}
\def\mn@eprint#1#2{\mn@eprint@#1:#2::\@nil}
\def\mn@eprint@arXiv#1{\href {http://arxiv.org/abs/#1} {{\tt arXiv:#1}}}
\def\mn@eprint@dblp#1{\href {http://dblp.uni-trier.de/rec/bibtex/#1.xml}
  {dblp:#1}}
\def\mn@eprint@#1:#2:#3:#4\@nil{\def\@tempa {#1}\def\@tempb {#2}\def\@tempc
  {#3}\ifx \@tempc \@empty \let \@tempc \@tempb \let \@tempb \@tempa \fi \ifx
  \@tempb \@empty \def\@tempb {arXiv}\fi \@ifundefined
  {mn@eprint@\@tempb}{\@tempb:\@tempc}{\expandafter \expandafter \csname
  mn@eprint@\@tempb\endcsname \expandafter{\@tempc}}}

\bibitem[\protect\citeauthoryear{{Baugh}}{{Baugh}}{2006}]{Baugh2006}
{Baugh} C.~M.,  2006, \mn@doi [Reports on Progress in Physics]
  {10.1088/0034-4885/69/12/R02}, 69, 3101

\bibitem[\protect\citeauthoryear{{Baugh} et~al.,}{{Baugh}
  et~al.}{2019}]{Baugh2019}
{Baugh} C.~M.,  et~al., 2019, \mn@doi [\mnras] {10.1093/mnras/sty3427}, \href
  {https://ui.adsabs.harvard.edu/abs/2019MNRAS.483.4922B} {483, 4922}

\bibitem[\protect\citeauthoryear{{Benson}}{{Benson}}{2010}]{Benson2010}
{Benson} A.~J.,  2010, \mn@doi [Physics Reports]
  {10.1016/j.physrep.2010.06.001}, 495, 33

\bibitem[\protect\citeauthoryear{{Berlind} \& {Weinberg}}{{Berlind} \&
  {Weinberg}}{2002}]{Berlind2003}
{Berlind} A.~A.,  {Weinberg} D.~H.,  2002, \mn@doi [\apj] {10.1086/341469},
  575, 587

\bibitem[\protect\citeauthoryear{{Bianchi} \& {Percival}}{{Bianchi} \&
  {Percival}}{2017}]{BianchiPercival2017}
{Bianchi} D.,  {Percival} W.~J.,  2017, \mn@doi [\mnras]
  {10.1093/mnras/stx2053}, 472, 1106

\bibitem[\protect\citeauthoryear{{Bower}, {Benson}, {Malbon}, {Helly}, {Frenk},
  {Baugh}, {Cole}  \& {Lacey}}{{Bower} et~al.}{2006}]{Bower2006}
{Bower} R.~G.,  {Benson} A.~J.,  {Malbon} R.,  {Helly} J.~C.,  {Frenk} C.~S.,
  {Baugh} C.~M.,  {Cole} S.,   {Lacey} C.~G.,  2006, \mn@doi [\mnras]
  {10.1111/j.1365-2966.2006.10519.x}, 370, 645

\bibitem[\protect\citeauthoryear{{Boylan-Kolchin}, {Springel}, {White},
  {Jenkins}  \& {Lemson}}{{Boylan-Kolchin} et~al.}{2009}]{BoylanKolchin2009}
{Boylan-Kolchin} M.,  {Springel} V.,  {White} S. D.~M.,  {Jenkins} A.,
  {Lemson} G.,  2009, \mn@doi [\mnras] {10.1111/j.1365-2966.2009.15191.x}, 398,
  1150

\bibitem[\protect\citeauthoryear{{Chandro-G{\'o}mez}
  et~al.,}{{Chandro-G{\'o}mez} et~al.}{2025}]{ChandroGomez2025}
{Chandro-G{\'o}mez} A.,  et~al., 2025, \mn@doi [\mnras]
  {10.1093/mnras/staf519}, 539, 776

\bibitem[\protect\citeauthoryear{{Cochran}}{{Cochran}}{1977}]{Cochran1977}
{Cochran} W.~G.,  1977, {Sampling Techniques}, 3rd edn.
John Wiley \& Sons, New York

\bibitem[\protect\citeauthoryear{{Cole}, {Lacey}, {Baugh}  \& {Frenk}}{{Cole}
  et~al.}{2000}]{Cole2000}
{Cole} S.,  {Lacey} C.~G.,  {Baugh} C.~M.,   {Frenk} C.~S.,  2000, \mn@doi
  [\mnras] {10.1046/j.1365-8711.2000.03879.x}, 319, 168

\bibitem[\protect\citeauthoryear{{Cooray} \& {Sheth}}{{Cooray} \&
  {Sheth}}{2002}]{Cooray2002}
{Cooray} A.,  {Sheth} R.,  2002, \mn@doi [Physics Reports]
  {10.1016/S0370-1573(02)00276-4}, 372, 1

\bibitem[\protect\citeauthoryear{{Cranmer}, {Brehmer}  \& {Louppe}}{{Cranmer}
  et~al.}{2020}]{Cranmer2020}
{Cranmer} K.,  {Brehmer} J.,   {Louppe} G.,  2020, \mn@doi [Proceedings of the
  National Academy of Sciences] {10.1073/pnas.1912789117}, 117, 30055

\bibitem[\protect\citeauthoryear{{DESI Collaboration}}{{DESI
  Collaboration}}{2016}]{DESI2016}
{DESI Collaboration} 2016, preprint (\mn@eprint {arXiv} {1611.00036})

\bibitem[\protect\citeauthoryear{{G{\'o}mez}, {Padilla}, {Helly}, {Lacey},
  {Baugh}  \& {Lagos}}{{G{\'o}mez} et~al.}{2022}]{Gomez:2022}
{G{\'o}mez} J.~S.,  {Padilla} N.~D.,  {Helly} J.~C.,  {Lacey} C.~G.,  {Baugh}
  C.~M.,   {Lagos} C.~D.~P.,  2022, \mn@doi [\mnras] {10.1093/mnras/stab3661},
  \href {https://ui.adsabs.harvard.edu/abs/2022MNRAS.510.5500G} {510, 5500}

\bibitem[\protect\citeauthoryear{{Gonzalez-Perez}, {Lacey}, {Baugh}, {Lagos},
  {Helly}, {Campbell}  \& {Mitchell}}{{Gonzalez-Perez}
  et~al.}{2014}]{GonzalezPerez2014}
{Gonzalez-Perez} V.,  {Lacey} C.~G.,  {Baugh} C.~M.,  {Lagos} C. d.~P.,
  {Helly} J.,  {Campbell} D. J.~R.,   {Mitchell} P.~D.,  2014, \mn@doi [\mnras]
  {10.1093/mnras/stt2410}, 439, 264

\bibitem[\protect\citeauthoryear{{Guo}, {White}, {Angulo}, {Henriques},
  {Lemson}, {Boylan-Kolchin}, {Thomas}  \& {Short}}{{Guo}
  et~al.}{2013}]{Guo2013}
{Guo} Q.,  {White} S.,  {Angulo} R.~E.,  {Henriques} B.,  {Lemson} G.,
  {Boylan-Kolchin} M.,  {Thomas} P.,   {Short} C.,  2013, \mn@doi [\mnras]
  {10.1093/mnras/sts115}, 428, 1351

\bibitem[\protect\citeauthoryear{Harris et~al.,}{Harris
  et~al.}{2020}]{harris2020numpy}
Harris C.~R.,  et~al., 2020, \mn@doi [Nature] {10.1038/s41586-020-2649-2}, 585,
  357

\bibitem[\protect\citeauthoryear{{Henriques}, {White}, {Thomas}, {Angulo},
  {Guo}, {Lemson}  \& {Springel}}{{Henriques} et~al.}{2013}]{Henriques2013}
{Henriques} B.~M.~B.,  {White} S.~D.~M.,  {Thomas} P.~A.,  {Angulo} R.~E.,
  {Guo} Q.,  {Lemson} G.,   {Springel} V.,  2013, \mn@doi [\mnras]
  {10.1093/mnras/stt415}, 431, 3373

\bibitem[\protect\citeauthoryear{{Horvitz} \& {Thompson}}{{Horvitz} \&
  {Thompson}}{1952}]{Horvitz1952}
{Horvitz} D.~G.,  {Thompson} D.~J.,  1952, \mn@doi [Journal of the American
  Statistical Association] {10.1080/01621459.1952.10483446}, 47, 663

\bibitem[\protect\citeauthoryear{Hunter}{Hunter}{2007}]{hunter2007matplotlib}
Hunter J.~D.,  2007, \mn@doi [Computing in Science \& Engineering]
  {10.1109/MCSE.2007.55}, 9, 90

\bibitem[\protect\citeauthoryear{{Jiang}, {Helly}, {Cole}  \& {Frenk}}{{Jiang}
  et~al.}{2014}]{Jiang2014}
{Jiang} L.,  {Helly} J.~C.,  {Cole} S.,   {Frenk} C.~S.,  2014, \mn@doi
  [\mnras] {10.1093/mnras/stu390}, 440, 2115

\bibitem[\protect\citeauthoryear{{Kaiser}}{{Kaiser}}{1986}]{Kaiser1986}
{Kaiser} N.,  1986, \mn@doi [\mnras] {10.1093/mnras/219.4.785}, 219, 785

\bibitem[\protect\citeauthoryear{{Kaiser}}{{Kaiser}}{1987}]{Kaiser1987}
{Kaiser} N.,  1987, \mn@doi [\mnras] {10.1093/mnras/227.1.1}, 227, 1

\bibitem[\protect\citeauthoryear{{Kennicutt}}{{Kennicutt}}{1983}]{Kennicutt1983}
{Kennicutt} Jr. R.~C.,  1983, \mn@doi [\apj] {10.1086/161261}, 272, 54

\bibitem[\protect\citeauthoryear{{Kugel} et~al.,}{{Kugel}
  et~al.}{2023}]{Kugel:2023}
{Kugel} R.,  et~al., 2023, \mn@doi [\mnras] {10.1093/mnras/stad2540}, \href
  {https://ui.adsabs.harvard.edu/abs/2023MNRAS.526.6103K} {526, 6103}

\bibitem[\protect\citeauthoryear{{Lacey} et~al.,}{{Lacey}
  et~al.}{2016}]{Lacey2016}
{Lacey} C.~G.,  et~al., 2016, \mn@doi [\mnras] {10.1093/mnras/stw1888}, 462,
  3854

\bibitem[\protect\citeauthoryear{{Landy} \& {Szalay}}{{Landy} \&
  {Szalay}}{1993}]{LandySzalay1993}
{Landy} S.~D.,  {Szalay} A.~S.,  1993, \mn@doi [\apj] {10.1086/172900}, 412, 64

\bibitem[\protect\citeauthoryear{{Laureijs} et~al.,}{{Laureijs}
  et~al.}{2011}]{Laureijs2011}
{Laureijs} R.,  et~al., 2011, preprint (\mn@eprint {arXiv} {1110.3193})

\bibitem[\protect\citeauthoryear{Matsakis \& Klock}{Matsakis \&
  Klock}{2014}]{matsakis2014rust}
Matsakis N.~D.,  Klock F.~S.,  2014, \mn@doi [ACM SIGAda Ada Letters]
  {10.1145/2692956.2663188}, 34, 103

\bibitem[\protect\citeauthoryear{{Merson} et~al.,}{{Merson}
  et~al.}{2013}]{Merson2013}
{Merson} A.~I.,  et~al., 2013, \mn@doi [\mnras] {10.1093/mnras/sts355}, 429,
  556

\bibitem[\protect\citeauthoryear{{Orsi}, {Lacey}, {Baugh}  \& {Infante}}{{Orsi}
  et~al.}{2008}]{Orsi:2008}
{Orsi} A.,  {Lacey} C.~G.,  {Baugh} C.~M.,   {Infante} L.,  2008, \mn@doi
  [\mnras] {10.1111/j.1365-2966.2008.14010.x}, 391, 1589

\bibitem[\protect\citeauthoryear{{Peebles}}{{Peebles}}{1980}]{Peebles1980}
{Peebles} P.~J.~E.,  1980, {The Large-Scale Structure of the Universe}.
Princeton University Press, Princeton

\bibitem[\protect\citeauthoryear{{Planck Collaboration}}{{Planck
  Collaboration}}{2014}]{Planck2013}
{Planck Collaboration} 2014, \mn@doi [\aap] {10.1051/0004-6361/201321591}, 571,
  A16

\bibitem[\protect\citeauthoryear{{S\'anchez} et~al.,}{{S\'anchez}
  et~al.}{2017}]{Sanchez2016}
{S\'anchez} A.~G.,  et~al., 2017, \mn@doi [\mnras] {10.1093/mnras/stw2443},
  464, 1640

\bibitem[\protect\citeauthoryear{{Schaye} et~al.,}{{Schaye}
  et~al.}{2023}]{Schaye:2023}
{Schaye} J.,  et~al., 2023, \mn@doi [\mnras] {10.1093/mnras/stad2419}, \href
  {https://ui.adsabs.harvard.edu/abs/2023MNRAS.526.4978S} {526, 4978}

\bibitem[\protect\citeauthoryear{{Sinha} \& {Garrison}}{{Sinha} \&
  {Garrison}}{2019}]{sinha2019corrfunc_avx}
{Sinha} M.,  {Garrison} L.,  2019, in Majumdar A.,  Arora R.,  eds, Software
  Challenges to Exascale Computing. Springer Singapore, pp 3--20,
  \mn@doi{10.1007/978-981-13-7729-7_1}

\bibitem[\protect\citeauthoryear{{Sinha} \& {Garrison}}{{Sinha} \&
  {Garrison}}{2020}]{sinha2020corrfunc}
{Sinha} M.,  {Garrison} L.~H.,  2020, \mn@doi [\mnras] {10.1093/mnras/stz3157},
  491, 3022

\bibitem[\protect\citeauthoryear{{Spergel} et~al.,}{{Spergel}
  et~al.}{2003}]{WMAP1}
{Spergel} D.~N.,  et~al., 2003, \mn@doi [\apjs] {10.1086/377226}, 148, 175

\bibitem[\protect\citeauthoryear{{Springel}, {White}, {Tormen}  \&
  {Kauffmann}}{{Springel} et~al.}{2001}]{Springel2001}
{Springel} V.,  {White} S. D.~M.,  {Tormen} G.,   {Kauffmann} G.,  2001,
  \mn@doi [\mnras] {10.1046/j.1365-8711.2001.04912.x}, 328, 726

\bibitem[\protect\citeauthoryear{{Springel} et~al.,}{{Springel}
  et~al.}{2005}]{Springel2005}
{Springel} V.,  et~al., 2005, \mn@doi [Nature] {10.1038/nature03597}, 435, 629

\bibitem[\protect\citeauthoryear{{Stiskalek}, {Perez}, {Genel}, {Somerville},
  {Angulo}  \& {Contreras}}{{Stiskalek} et~al.}{2026}]{Stiskalek2026}
{Stiskalek} R.,  {Perez} L.~A.,  {Genel} S.,  {Somerville} R.~S.,  {Angulo}
  R.~E.,   {Contreras} S.,  2026, \mn@doi [\apj] {10.3847/1538-4357/ae878b},
  1007, 112

\bibitem[\protect\citeauthoryear{{\noopsort{Van Daalen}}{van Daalen},
  {Henriques}, {Angulo}  \& {White}}{{\noopsort{Van Daalen}}{van Daalen}
  et~al.}{2016}]{vanDaalen2016}
{\noopsort{Van Daalen}}{van Daalen} M.~P.,  {Henriques} B. M.~B.,  {Angulo}
  R.~E.,   {White} S. D.~M.,  2016, \mn@doi [\mnras] {10.1093/mnras/stw405},
  \href {https://ui.adsabs.harvard.edu/abs/2016MNRAS.458..934V} {458, 934}

\bibitem[\protect\citeauthoryear{{\noopsort{Van Rossum}}{van Rossum} \&
  Drake}{{\noopsort{Van Rossum}}{van Rossum} \&
  Drake}{2009}]{vanrossum2009python}
{\noopsort{Van Rossum}}{van Rossum} G.,  Drake F.~L.,  2009, {Python 3
  Reference Manual}.
CreateSpace, Scotts Valley, CA

\bibitem[\protect\citeauthoryear{Virtanen et~al.,}{Virtanen
  et~al.}{2020}]{virtanen2020scipy}
Virtanen P.,  et~al., 2020, \mn@doi [Nature Methods]
  {10.1038/s41592-019-0686-2}, 17, 261

\bibitem[\protect\citeauthoryear{{Wechsler} \& {Tinker}}{{Wechsler} \&
  {Tinker}}{2018}]{Wechsler2018}
{Wechsler} R.~H.,  {Tinker} J.~L.,  2018, \mn@doi [\araa]
  {10.1146/annurev-astro-081817-051756}, \href
  {https://ui.adsabs.harvard.edu/abs/2018ARA&A..56..435W} {56, 435}

\bibitem[\protect\citeauthoryear{{Zheng} et~al.,}{{Zheng}
  et~al.}{2005}]{Zheng2005}
{Zheng} Z.,  et~al., 2005, \mn@doi [\apj] {10.1086/466510}, 633, 791

\makeatother
\end{thebibliography}

\appendix

\section{Closed-form partition-selection variance}
\label{app:closed-form-variance}

The scatter of the estimator from one selection to another can be computed without repeating the selection, and this confirms (Fig.~A1) that the empirical variance fitted in Section~5.5 is what the random choice of partitions predicts, rather than a coincidence of the samples tested. The calculation uses one property of the selection: a pair survives only if every partition it involves is selected.

Write $S$ for the set of $m$ selected partitions, $DD_{aa}$ for the pair count within partition $a$, and $DD_{ab}$ for the count between partitions $a$ and $b$, all within a fixed separation bin. The two terms of equation~\eqref{eq:corrected-count} are then sums over $S$,

\begin{equation}
DD_{\rm auto} = \sum_{a \in S} DD_{aa}, \qquad
DD_{\rm cross} = \sum_{\substack{a<b\\ a,b\, \in\, S}} DD_{ab},
\end{equation}

and $DD_{\rm auto}^{(k)}$, $DD_{\rm cross}^{(k)}$ are the same sums taken over all $k$ partitions. The corrected count is $DD = w_{\rm auto}\,DD_{\rm auto} + w_{\rm cross}\,DD_{\rm cross}$, equation~\eqref{eq:corrected-count}. The pair counts are fixed by the catalogue and the weights by $m$ and $k$; all of the randomness is in $S$, so every $\mathbb{E}[\cdot]$ below is an average over the choice of $S$, as in Section~\ref{sec:sauce-derivation}.

Extending equation~\eqref{eq:survival}, the probability that $j$ specified partitions are all selected is

\begin{equation}
p_j = \frac{m(m-1)\cdots(m-j+1)}{k(k-1)\cdots(k-j+1)} .
\end{equation}

A term of $DD_{\rm auto}$ names one partition and survives with probability $p_1$; a term of $DD_{\rm cross}$ names two and survives with probability $p_2$. Taking expectations term by term returns equations~\eqref{eq:exact-auto} and \eqref{eq:exact-cross}.

The variance needs the same rule applied to products. Squaring $DD$ multiplies every term by every other, and such a product survives only if all the partitions the two terms name between them are selected, so its expectation is $p_j$ with $j$ the number of distinct partitions in that union. Four sums over the full catalogue, with $DD_{ba} \equiv DD_{ab}$, are enough to write the moment down, one for each way two terms can overlap, by coinciding or by sharing an index,

\begin{equation}
\begin{aligned}
T_1 &= \sum_a DD_{aa}^2, &\qquad
T_2 &= \sum_a DD_{aa} \sum_{b \neq a} DD_{ab}, \\
T_3 &= \sum_{a<b} DD_{ab}^2, &\qquad
T_4 &= \sum_a \Big( \sum_{b \neq a} DD_{ab} \Big)^{\!2}.
\end{aligned}
\label{eq:coincidence-sums}
\end{equation}

Expanding the square, two auto terms name one partition if they coincide, contributing $T_1$ at $p_1$, and two if they do not, contributing the remainder $(DD_{\rm auto}^{(k)})^2 - T_1$ at $p_2$. An auto and a cross term name two if the auto index lies in the cross pair, contributing $T_2$, and three if it does not, contributing $DD_{\rm auto}^{(k)} DD_{\rm cross}^{(k)} - T_2$. Two cross terms name two, three or four: their pairs coincide, giving $T_3$; share one index, giving $T_4 - 2T_3$, the subtraction removing the coincident products $T_4$ also collects, each counted from both ends; or are disjoint, giving the remainder $(DD_{\rm cross}^{(k)})^2 - T_4 + T_3$. Hence

\begin{align}
\mathbb{E}\big[DD^2\big] = {}&
w_{\rm auto}^2 \Big[ p_1 T_1 + p_2 \big( (DD_{\rm auto}^{(k)})^2 - T_1 \big) \Big] \nonumber \\
&{}+ 2\,w_{\rm auto}w_{\rm cross} \Big[ p_2 T_2 + p_3 \big( DD_{\rm auto}^{(k)} DD_{\rm cross}^{(k)} - T_2 \big) \Big] \nonumber \\
&{}+ w_{\rm cross}^2 \Big[ p_2 T_3 + p_3 \big( T_4 - 2 T_3 \big) \nonumber \\
&\qquad\quad {}+ p_4 \big( (DD_{\rm cross}^{(k)})^2 - T_4 + T_3 \big) \Big] ,
\label{eq:second-moment}
\end{align}

and $\mathrm{Var}[DD] = \mathbb{E}[DD^2] - \mathbb{E}[DD]^2$. Every sum on the right is a property of the full $k$-partition catalogue; the selection enters only through $w_{\rm auto}$, $w_{\rm cross}$ and $p_j$, each a function of $m$ and $k$ alone.

We validated equation~(A4) against brute-force evaluation on all $\binom{8}{m}$ possible selections at $k = 8$, recovering the enumerated variance to machine precision at every $m$ from 2 to 7.

Brute-force enumeration is only feasible for small $k$, so for the production partitioning we evaluate equation~\eqref{eq:second-moment} directly on the measured pair counts, with no averaging over random selections.

\subsection{The random normalisation}
\label{app:ratio-variance}

$\mathrm{Var}[DD]$ is not by itself the variance of the estimator. The quantity measured is $\hat\xi = DD/RR - 1$, and $RR$ is built from the galaxy count actually selected, $N_{\rm sel} = \sum_{a \in S} n_a$ with $n_a$ the number of galaxies in partition $a$: $RR \propto W \equiv N_{\rm sel}(N_{\rm sel}-1)$, the number of ordered pairs of selected galaxies, with a constant that does not depend on the selection and cancels below. $RR$ is therefore random too, and correlated with $DD$, since a selection that happens to contain denser-than-average partitions raises both.

\begin{figure}
	\centering
	\includegraphics[width=0.75\columnwidth]{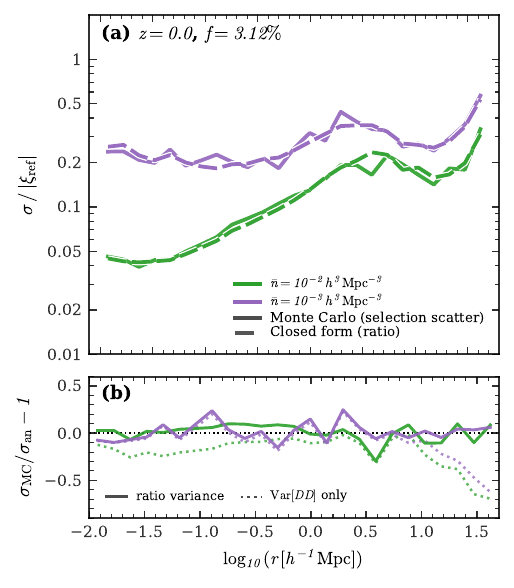}
	\caption{Monte Carlo versus closed-form partition-selection variance against pair separation, at $z=0$ on the P-Millennium; the other two redshifts agree equally well (not shown). \textit{(a)}: empirical $\sigma/|\xi_{\rm ref}|(r)$ at $m=32$ from $205$ and $105$ independent selections for the denser and sparser threshold respectively (solid, coloured by number-density threshold), with the closed-form ratio variance of equation~(\ref{eq:delta-method}) overplotted (dashed). \textit{(b)}: the residual $\sigma_{\rm MC}/\sigma_{\rm an} - 1$ in each bin, where $\sigma_{\rm MC}$ is the empirical scatter of panel (a) and $\sigma_{\rm an}$ one of two closed-form predictions.
    \textit{Solid}: the ratio variance of equation~(\ref{eq:delta-method}). \textit{Dotted}: the same calculation with the normalisation held fixed at its expectation, $\sqrt{\mathrm{Var}[DD]}/\mathbb{E}[RR]$; the gap between the two curves is the variance that cancels between numerator and denominator. The sparsest threshold, $\bar n = 10^{-4}\,h^3\,\mathrm{Mpc}^{-3}$, is shot-noise limited and is excluded here as it is from the P-Millennium-only recipe fit (Section~\ref{sec:variance-recipe}).}
	\label{fig:variance-validation}
\end{figure}

Because $W$ has the same form as $DD$, equation~\eqref{eq:second-moment} applies to it directly. Multiplying out $N_{\rm sel}(N_{\rm sel}-1)$,
\begin{equation}
W = \sum_{a \in S} \left(n_a^2 - n_a\right) + \sum_{\substack{a<b\\ a,b\, \in\, S}} 2 n_a n_b ,
\end{equation}
again a sum over selected partitions plus a sum over selected pairs of them.

Replacing $w_{\rm auto} DD_{aa}$ by $n_a^2 - n_a$ and $w_{\rm cross} DD_{ab}$ by $2 n_a n_b$ in equation~(A4) therefore gives $\mathbb{E}[W^2]$. The bilinear form of the same expansion, in which each square becomes a product of a $DD$ term and a $W$ term and the auto--cross term is split into its two orderings, gives $\mathbb{E}[DD\,W]$.
With these, the delta method gives the variance of the estimator,

\begin{equation}
\begin{split}
\mathrm{Var}[\hat\xi] \simeq \left(\frac{\mathbb{E}[DD]}{\mathbb{E}[RR]}\right)^{\!2}
\biggl[ &\frac{\mathrm{Var}[DD]}{\mathbb{E}[DD]^2}
+ \frac{\mathrm{Var}[W]}{\mathbb{E}[W]^2} \\
&- \frac{2\,\mathrm{Cov}[DD, W]}{\mathbb{E}[DD]\,\mathbb{E}[W]} \biggr].
\end{split}
\label{eq:delta-method}
\end{equation}

The three moments are exact, and agree with enumeration to the same precision as above; only truncating the ratio expansion is approximate. That truncation error is set by the partition-to-partition scatter in galaxy count, measured by its coefficient of variation $c_n$: the standard deviation of $n_a$ over the $k$ partitions divided by their mean.

Enumeration at $k = 8$, with partitions given the $c_n$ of our samples,
puts it at about one per cent, never more than 1.4 per cent, always in
the sense of underestimating $\sigma$.

The bracket shows whether dividing by $RR$ raises or lowers the fractional variance. Write $c_{DD}$ and $c_W$ for the coefficients of variation of $DD$ and $W$ across selections, and $\rho_{DW}$ for their correlation. The bracket is then $c_{DD}^2 + c_W^2 - 2\rho_{DW}\,c_{DD}\,c_W$, which is smaller than the numerator's own $c_{DD}^2$ when $c_W < 2\rho_{DW}\,c_{DD}$. Here $c_W \simeq 2 c_n \sqrt{(1-f)/m}$, which reproduces the enumerated $c_W$ to 2--4 per cent; $\mathrm{Var}[W]/\mathbb{E}[W]^2$ therefore falls as $(1-f)/m$.

Our samples satisfy $c_W < 2\rho_{DW} c_{DD}$ at every separation and
for both densities.

Beyond $r \approx 5\,h^{-1}\,\mathrm{Mpc}$ almost every pair spans two partitions, so $DD$ and $W$ both scale as the square of the selected galaxy count and are nearly perfectly correlated, $\rho_{DW} = 0.96$ in the largest bin of Fig.~\ref{fig:variance-validation}: a selection that catches denser partitions raises numerator and denominator together, and the fluctuation cancels from $\hat\xi$. The bracket falls from ${\approx}0.9\,c_{DD}^2$ at $r \sim 5\,h^{-1}\,\mathrm{Mpc}$ to $0.08\,c_{DD}^2$ in the largest bin. Holding the normalisation fixed at its expectation ignores this cancellation, and Fig.~\ref{fig:variance-validation} shows the cost: it overstates the scatter by up to a factor of about three in the largest bins, whereas equation~\eqref{eq:delta-method} matches it to within a few per cent in the median.
\section{Galaxy formation models}
\label{app:galaxy-formation-models}

The two \texttt{GALFORM} variants were calibrated for different
purposes: L16 \citep{Lacey2016} to reproduce galaxy populations from
the ultraviolet to the submillimetre from $z = 0$ to $z = 6$,
and GP14 \citep{GonzalezPerez2014} to reproduce the local $b_{\rm J}$- and
$K$-band luminosity functions with a universal solar-neighbourhood stellar initial mass function (IMF). They differ in the
parameters of Table~B1, which control supernova feedback and the return
of ejected gas, AGN feedback, star formation and the IMF.
L16 forms stars in bursts with a top-heavy IMF, which helps it reproduce the number counts of submillimetre galaxies at high redshift; GP14 uses the \citet{Kennicutt1983} IMF throughout. These differences change the stellar masses of the galaxies in each halo, how many galaxies populate each halo and hence which galaxies enter a sample of given number density. For $\bar{n} = 0.0408\,h^{3}\,{\rm Mpc}^{-3}$ at $z = 0$ in the P-Millennium, satellites (non-central galaxies) make up 39.6 per cent of the L16 sample and 39.9 per cent of the GP14 sample. The full-box correlation functions of the two samples differ by up to 8 per cent, L16 being lower at all but the smallest separations and by about 2 per cent on the largest scales measured for these samples, $r \approx 9\,h^{-1}$\,Mpc; at the same density at $z \approx 1.5$ they differ by less than 3 per cent.

\begin{table}
\caption{Parameters that differ between the two \texttt{GALFORM}
variants; all others are identical. Values are those recalibrated for
the P-Millennium by \citet{Baugh2019} where recalibration changed them,
and as published otherwise. The recalibration changed
$\alpha_{\rm cool}$ (from 0.60) and $V_{\rm hot}$ (from
425\,km\,s$^{-1}$) in GP14, and $\alpha_{\rm hot}$ (from 3.2) and
$\alpha_{\rm reheat}$ (from 0.64) in L16, to the values listed.
$V_{\rm hot}$ applies to both quiescent and burst star formation. Both
variants use a \citet{Kennicutt1983} IMF for quiescent star formation;
in bursts L16 uses a top-heavy IMF with slope $x = 1$.}
\label{tab:galform}
\begin{tabular}{@{}l >{\raggedright\arraybackslash}p{3.2cm} c c@{}}
\hline
Parameter & Description & L16 & GP14 \\
\hline
\multicolumn{4}{@{}l}{\textit{Supernova feedback and gas return}} \\
$V_{\rm hot}$ & Mass-loading velocity scale (km\,s$^{-1}$) & 320 & 380 \\
$\alpha_{\rm hot}$ & Mass-loading power-law slope & 3.4 & 3.2 \\
$\alpha_{\rm reheat}$ & Return rate of ejected gas & 1.00 & 1.26 \\
\multicolumn{4}{@{}l}{\textit{AGN feedback}} \\
$\epsilon_{\rm AGN}$ & Fraction of Eddington luminosity for heating & 0.01 & 0.04 \\
$\alpha_{\rm cool}$ & Hot-halo threshold for AGN feedback & 0.80 & 0.72 \\
\multicolumn{4}{@{}l}{\textit{Star formation}} \\
$\nu_{\rm SF}$ & H$_2$ depletion rate (Gyr$^{-1}$) & 0.74 & 0.50 \\
$f_{\rm stab}$ & Disc stability threshold & 0.90 & 0.80 \\
IMF in bursts & & Top-heavy & Kennicutt \\
\hline
\end{tabular}
\end{table}
\section{Shape of the partition-selection distribution}\label{app:variance-skew}

\begin{figure*}
	\centering
	\begin{minipage}[t]{0.33\textwidth}
		\includegraphics[width=\textwidth]{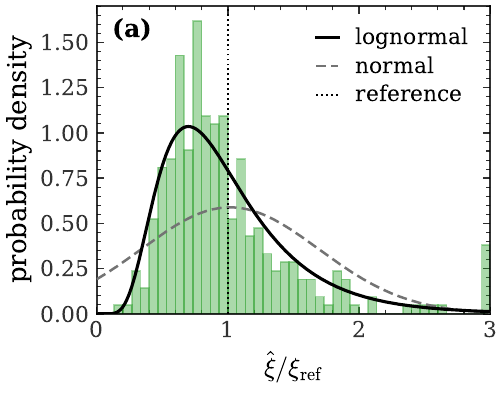}
	\end{minipage}\hfill
	\begin{minipage}[t]{0.33\textwidth}
		\includegraphics[width=\textwidth]{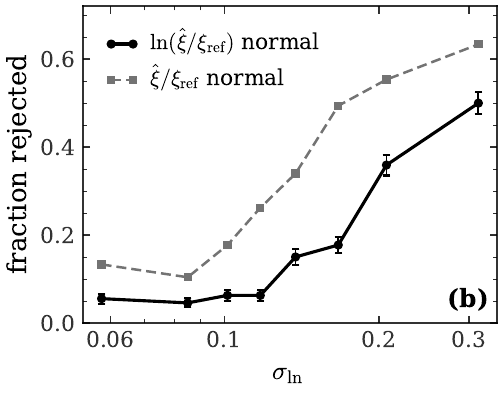}
	\end{minipage}\hfill
	\begin{minipage}[t]{0.33\textwidth}
		\includegraphics[width=\textwidth]{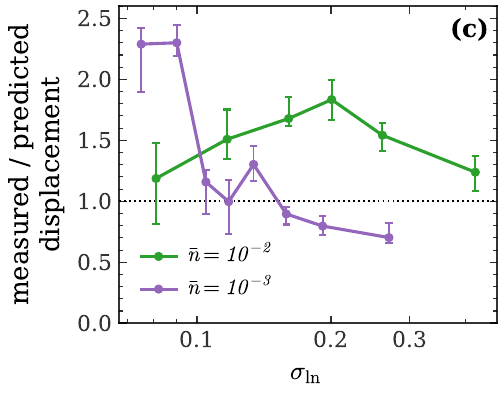}
	\end{minipage}
	\caption{The distribution of $R = \hat\xi/\xi_{\rm ref}$ over selections. \textit{(a)}: one cell ($\bar n = 10^{-2}\,h^3\,\mathrm{Mpc}^{-3}$, $m=8$, $r = 7.6\,h^{-1}\,\mathrm{Mpc}$, $315$ selections), with a lognormal (solid) and a normal (dashed) given the sample mean and standard deviation of $\ln R$ and of $R$ respectively; the rightmost bin collects every selection with $R > 3$. \textit{(b)}: fraction of the $3294$ uncensored cells in which the Anderson--Darling test rejects normality at five per cent, for $\ln R$ (solid) and $R$ (dashed), against $\sigma_{\ln}$, the standard deviation of $\ln R$ over a cell's selections, with binomial errors. \textit{(c)}: the measured displacement of the median below the reference, $1 - {\rm med}$, divided by the lognormal prediction $1 - e^{-\sigma_{\ln}^2/2}$ of equation~\eqref{eq:lognormal-moments}, against $\sigma_{\ln}$; the dotted line marks unity. Each point is the ratio of the binned medians of the two quantities, in eight equal-count bins of $\sigma_{\ln}$ per sample; bars are the 16th--84th percentiles under bootstrap resampling of the cells in the bin, and are a lower bound. Bins below $\sigma_{\ln} \approx 0.07$ are not drawn (Appendix~\ref{app:mean-vs-median}).}
	\label{fig:variance-lognormal}
\end{figure*}

Appendix~A gives the width of the scatter between selections; this
appendix gives its shape, and justifies three choices made in
Section~4. The scatter is close to lognormal (Appendix~C1), so the
$1\sigma$ band is stated in $\log_{10} R$. Because the distribution is
skewed, a typical selection falls below the reference even though the
mean does not (Appendix~C2), so the bias is reported as a mean. And
discarding selections that return $\hat{\xi} \le 0$ biases the result
upwards (Appendix~C3), so the acceptance criterion counts them as
failures.

Unless stated otherwise we use the P-Millennium L16 catalogue at
$z = 0$ and the samples $\bar{n} = 10^{-2}$ and
$10^{-3}\,h^{3}\,{\rm Mpc}^{-3}$.
We use the 25 bins of the coarse grid of Section~4.2 with centres below
$50\,h^{-1}$\,Mpc.
We call each $(r, m, \bar{n})$ combination a cell and study the ratio $R$ of
equation~(11) over its selections: 105 in 95 per cent of uncensored
cells, and between 192 and 415 in the rest.
A cell in which any selection returns $\hat{\xi} \le 0$ is censored:
$\ln R$ is undefined for that selection, and the missing low tail would
distort any test of shape. Of the 3425 cells with $2 \le m < k$
and at least 100 selections, 131 are censored; Appendices~C1 and~C2
use only the 3294 uncensored cells, and
Appendix~C3 treats the censored ones. We measure the width of a cell by
$\sigma_{\ln}$, the standard deviation of $\ln R$ over its selections,
which spans 0.008 to 1.07.

\subsection{Normality of the log-ratio}\label{app:lognormal-test}

Fig.~\ref{fig:variance-lognormal}(a) shows one of the widest cells, $\sigma_{\ln} = 0.49$, with a lognormal and a normal overplotted, each given the moments of the data rather than fitted to the histogram. The lognormal better matches the floor at $\hat\xi = 0$ and the long upper tail, though it still understates the skewness. Fig.~\ref{fig:variance-lognormal}(b) extends the comparison to every cell, applying the Anderson--Darling test for normality to $\ln R$ and to $R$.

The log-ratio $\ln R$ is closer to normal than $R$: its Anderson--Darling statistic is smaller in $73$ per cent of cells, and normality is rejected at the five per cent level for $\ln R$ in $18$ per cent of cells against $34$ per cent for $R$, with the lower rate for $\ln R$ at every $\sigma_{\ln}$. For the $986$ cells with $\sigma_{\ln} < 0.10$ the rejection rate for $\ln R$ is $4.9$ per cent, the rate expected if every cell were exactly lognormal. In the widest sixth of cells, $\sigma_{\ln} \geq 0.22$, it rises to $47$ per cent: the denser sample has a heavier tail than lognormal, and the sparser one tends back towards a Gaussian as shot noise dominates.
The contrast holds at $z \approx 1.5$, where normality is rejected in
14 per cent of the 3836 uncensored cells for $\ln R$ and $22$ per cent for $R$. Neighbouring cells share partitions and pairs, so these rates describe the ensemble rather than forming independent tests.

The acceptance criterion is met at $\sigma_{\ln} \lesssim 0.12$, at the edge of the range $\sigma_{\ln} < 0.10$ in which the test finds no departure from lognormality, so the $1\sigma$ band can be stated in $\log_{10}R$ and its width predicted from the variance.

\subsection{Skew and the mean--median offset}\label{app:mean-vs-median}

Equations~(4) and~(5) make $\mathbb{E}[DD]$ equal to $(m/k)^2$ times
the full-box count at every $m$, matching the normalisation of $RR$ on
average, so $\hat{\xi}$ is unbiased to leading order (Section~3.3).
We test this in every uncensored cell by dividing the offset of the mean of $R$ from unity by its standard error, giving $t$, which for an unbiased estimator follows a unit normal. The median $|t|$ is $0.71$ against the $0.67$ expected, the $68$th percentile is $1.07$ against $1.00$,
and 6 per cent of cells lie beyond $|t| = 2$, against 4.6 per cent.
The typical mean offset, the median over cells, is $-0.0005$. The cells are not independent, so this is a consistency check rather than a formal test, but it excludes a bias at the per-cent level.

\begin{figure}
	\centering
	\includegraphics[width=0.7\columnwidth]{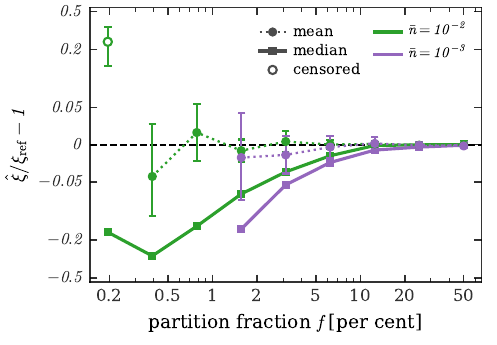}
\caption{Mean (dotted, circles, with standard errors) and median
(solid, squares) of $R - 1$ against partition fraction $f$ at
$r = 7.6\,h^{-1}\,\mathrm{Mpc}$, for both samples, at
$m = 2, 4, 8, \ldots$ with at least 100 selections returning
$\hat\xi > 0$. Fig.~\ref{fig:variance-lognormal}(a) shows the
distribution at one point of the green curve. The open circle marks a
cell in which more than five per cent of selections return
$\hat\xi \le 0$, biasing both statistics
(Appendix~\ref{app:positivity-censoring}). The $y$-axis is linear
within $\pm 0.05$ and logarithmic outside it.}
	\label{fig:variance-skew}
\end{figure}

The median is biased low, as Fig.~\ref{fig:variance-skew} shows at $r = 7.6\,h^{-1}\,\mathrm{Mpc}$: the mean stays at zero at every uncensored $f$, while the median lies below it and approaches zero only as $f$ grows. Across all uncensored cells the median offset, ${\rm med}(R) - 1$, is typically $-0.010$, more than an order of magnitude larger than the mean offset, consistently negative, and larger at smaller $f$.

The offset follows from the lognormal shape. For a lognormal variable of unit mean whose logarithm has standard deviation $\sigma_{\ln}$, the median, and the relation between $\sigma_{\ln}$ and the coefficient of variation $c$, are
\begin{equation}
{\rm med} = e^{-\sigma_{\ln}^2/2}, \qquad
\sigma_{\ln}^2 = \ln\!\left(1 + c^2\right).
\label{eq:lognormal-moments}
\end{equation}
For $R$, $c$ is the fractional scatter $\sigma/|\xi_{\rm ref}|$ of equation~\eqref{eq:delta-method}, so for a lognormal distribution the median offset is set by the variance. Fig.~\ref{fig:variance-lognormal}(c) tests this, comparing binned medians of the measured and predicted $1 - {\rm med}$, since a per-cell ratio is too noisy at small $\sigma_{\ln}$. With no free parameter, the prediction follows the measurement over more than a decade to within a factor of about two; it understates the offset at $\bar n = 10^{-2}\,h^3\,\mathrm{Mpc}^{-3}$ and overstates it in the widest cells at $10^{-3}\,h^3\,\mathrm{Mpc}^{-3}$.
For a single cell the predicted offset, about $\sigma_{\ln}^2/2$, is
smaller than the sampling error of a median over 105 selections,
about $0.12\,\sigma_{\ln}$, up to $\sigma_{\ln} \approx 0.25$; binning
the cells reduces that error, and below $\sigma_{\ln} \approx 0.07$
even the binned medians cannot measure the offset, so panel~(c) starts
there.

A median of $R$, or a mean of $\log_{10}R$, is therefore offset by $\sigma_{\ln}^2/(2\ln 10) \approx 0.217\,\sigma_{\ln}^2$ dex even for an exactly unbiased estimator: a tenth of the $0.05$ dex tolerance at $\sigma_{\ln} = 0.15$, and all of it at $\sigma_{\ln} = 0.5$. At small $f$ either statistic would mostly measure the skew of the distribution, so we report the bias as a mean of $R$.

\subsection{Positivity censoring in the sparse limit}\label{app:positivity-censoring}

At $m = 2$, 11 and 32 per cent of selections and bins over
${0.01 \le r < 50\,h^{-1}}$\,Mpc return $\hat{\xi} \le 0$ for the denser and sparser samples respectively.
At small separations, where $\xi \gg 1$, this happens only when the selected partitions contain no pair in the bin. Discarding these selections keeps only those that contained a pair, removing the low tail and biasing both the mean and the median upwards.

The bias is confined to a few cells but is large in them. In the $17$ cells with more than five per cent of selections returning $\hat\xi \leq 0$ and at least 100 returning $\hat\xi > 0$, the median over cells of the mean offset of the remaining selections, $\langle R \rangle - 1$, is $+0.351$, against $-0.0005$ for the uncensored cells, and it reaches $+3.5$; almost all are positive. The bias tracks the censored fraction: for $\bar n = 10^{-3}\,h^3\,\mathrm{Mpc}^{-3}$ at $r = 0.023\,h^{-1}\,\mathrm{Mpc}$, the fraction is $34.3$, $5.7$ and $0.0$ per cent at $m = 2$, $4$ and $8$, and the median offset, ${\rm med}(R) - 1$, falls from $+0.32$ to a value consistent with zero.

\bsp
\label{lastpage}
\end{document}